\documentstyle[11pt]{article}
\begin{document}
\parskip 5pt plus 1pc
\parindent=16pt
\parskip 5pt plus 1pc
\parindent=16pt
\thispagestyle{empty}
\newcommand{\lsa}{Lie superalgebra}
\newcommand{\sa}{superalgebra}
\newcommand{\lsas}{Lie superalgebras}
\newcommand{\sas}{superalgebras}
\newcommand{\beq}{\begin{eqnarray}}
\newcommand{\eeq}{\end{eqnarray}}
\newcommand{\beqn}{\begin{equation}}
\newcommand{\eeqn}{\end{equation}}
\newcommand{\np}[4]{Nucl.\ Phys.\ #1{\bf #2} (#4) #3}
\newcommand{\cmp}[3]{Commun.\ Math.\ Phys.\ {\bf #1} (#3) #2}
\newcommand{\pl}[4]{Phys.\ Lett.\ #1{\bf #2} (#4) #3}
\newcommand{\af}{\alpha}
\newcommand{\bt}{\beta}
\newcommand{\r}{\gamma}
\newcommand{\p}{\rho}
\newcommand{\k}{\kappa}
\newcommand{\0}{\theta}
\newcommand{\sg}{\sigma}
\newcommand{\h}[1]{H^ {\frac {\infty}{2} +#1}}
\newcommand{\vt}{\vartheta}
\newcommand{\df}{\partial}
\newcommand{\td}{\tilde}
\newcommand{\os}{\widehat{OSP(1, 2)}}
\newcommand{\OSP}{OSP(1, 2)}
\newcommand{\rar}{\rightarrow}

\newtheorem{pp}{Proposition}
\newtheorem{tm}{Theorem}
\newtheorem{lm}{Lemma}
\newtheorem{corollary}{Corollary}
\newcommand{\bpp}{\begin{pp}}
\newcommand{\epp}{\end{pp}}
\newcommand{\btm}{\begin{tm}}
\newcommand{\etm}{\end{tm}}
\newcommand{\blm}{\begin{lm}}
\newcommand{\elm}{\end{lm}}
\newcommand{\bc}{\begin{corollary}}
\newcommand{\ec}{\end{corollary}}
\newcommand{\thf}[3]{\vartheta \left[ \begin{array}{c} #1\\0 \end{array}
\right]
(#2,#3)}
\newcommand{\no}{\nonumber}
\newcommand{\script}{\scriptstyle}
\newcommand{\scriptscript}{\scriptscriptstyle}
\newcommand{\crossst}{\searrow \hspace{-1em}\swarrow}
\newcommand{\crossnd}{\nearrow \hspace{-1em}\nwarrow}
\newcommand{\crossrd}{\searrow \hspace{-1em}\nearrow}
\newcommand{\crossth}{\nwarrow \hspace{-1em}\swarrow}
\begin{flushright}
{\large AS-ITP-93-22} \\
Revised Version\\
June 1993
\end{flushright}
\vspace{10ex}
\centerline{\Large $G/G$ Gauged Supergroup Valued WZNW Field Theory}
\vspace{10ex}
\centerline{\large {\sc Jiang-Bei Fan}$^{b}$
and {\sc Ming YU}$^{a,b}$}
\vspace{2ex}
\centerline{\it $^a$CCAST (World Laboratory)}
\centerline{\it P.O.Box 8730, Beijing 100080, P.R.China}
\vspace{1ex}
\centerline{\it $^b$Institute of Theoretical Physics, Academia Sinica}
\centerline{\it P.O.Box 2735, Beijing 100080, P.R.China}
\vspace{6ex}
\centerline{\large \it Abstract}
\vspace{1ex}
\begin{center}
\begin{minipage}{130mm}
The $G/G$ gauged supergroup valued WZNW theory is considered. It is shown
that for $G=\OSP$, the $G/G$ theory tensoring a ($b$, $c$, $\beta$, $\gamma$)
system is equivalent to the non-critical fermionic theory.
The relation between integral or half integral moded affine superalgebra
and its reduced theory, the NS or R
superconformal algebra, is discussed in detail. The physical state space,
i.e. the BRST semi-infinite cohomology, is calculated, for the $\OSP/\OSP$
theory.
\end{minipage}
\end{center}
\vfill
\hrule
\newpage
\section{Introduction}
\pagenumbering{arabic}
In recent years, $2D$ gravity and matrix models have aroused much interest
of the string theorists \cite{BK,DS,GM1,GM2,LZ3,LZ4,W2,BMP1}.
 The goal is to find (non-perturbative) solutions
to the string theories in space-time dimensions other than the critical ones.
A striking feature of the non-critical string theory is the appearance of
infinitely many copies of the physical states with non-standard ghost numbers.
More recently, it has been found that the combination of the matter sector, the
Liouville sector and the reparametrization ghost
gives a $2D$ topological field theory, which is equivalent
to the $SL(2,R)/SL(2,R)$ gauged WZNW model \cite{AGS,HY1,HY2}.
The key observation is that
by adding a new term, $\partial J^{Tot,3}$, to the total energy momentum
tensor, $T^{Tot}(z)$, of the $SL(2,R)/SL(2,R)$ theory, one obtains the matter
and Liouville part of the
non-critical strings in the form of the Hamiltonian reduction of the WZNW
model.
The Liouville part is essentially the remainder of the gauge field.
The conformal dimensions of the ghosts also get re-adjusted when twisted
by $\partial J^{Tot,3}$ term.

There are some generalizations of the above construction. One way is to
consider
W-gravity coupled to W-matter, and naturally one would expect a
$SL(N,R)/SL(N,R)$ model \cite{ASY,S} ($W_N$ string), or a general $G/G$ theory.
Another way, which is the main subject of our present paper, is to look into
the non-critical fermionic string and find out its corresponding topological
field theory.
With respect to the latter approach,
much of recent work has been focused on the Hamiltonian reduction of the
super-group valued WZNW theory which gives rise to a super-Toda and (extended)
superconformal field theory \cite{II,Z,DGN,KMN,EH,IMP1,IMP2,I}.
However, the physical state
space of the non-critical fermionic string theory has not been worked out
completely (see ref.\cite{LZ5} for a remark on this point).
In this paper, we shall show that when
gluing together the super-Liouville, super-conformal matter, and the
super-reparametrization ghost,
one obtains
a topological super-conformal field theory, which is essentially
a $G/G$ model with $G=\OSP $.
The physical states of the $OSP(1,2)/OSP(1,2)$ theory, identified with
the BRST non-trivial states, is believed to be in one to one correspondence
with that of the non-critical fermionic string.
It should be  possible to apply our
method to the more generally extended superconformal field theory
\cite{DPYZ,BFK}
coupled to the generalized supergravities.

The essential ingredients of the WZNW theory is encoded in its current
algebra, the Kac-Moody algebra. It is also clear that it is the structure of
the
algebra modules that determine the physical states in $G/G$ WZNW theory.
The general structure of the affine Kac-Moody modules has been extensively
studied \cite{KK,K3,FFr,FF}. In our previous paper\cite{FY1}, some
results concerning the affine Kac-Moody modules have been generalized to the
case of affine Kac-Moody superalgebras. In this paper, we try to solve a
remaining issue, namely, the BRST semi-infinite cohomology of the $G/G$
WZNW theory with $G$ in general a supergroup,
which might be relevant to our understanding of the (gauged)supergroup
valued WZNW theory, as well as the role of 2d supergravity.

It is known that under Hamiltonian reduction, the representation spaces of
the $\widehat{SL(N)}$ algebra
(resp.\ $\widehat{OSP(N, 2)}$) are equivalent to that of the
$W_{N}$ algebra (resp.\ $N$-extended superconformal field theory) by imposing
constraints on the currents \cite{BO1,BO}.
 It may be the essential reason why the structure
of the Virasoro module is similar to that of the $\widehat{SL(2)}$ module
\cite{FFu,FF}.
{}From this point of view,  we can understand the correspondence between BRST
states of $2D$ gravity and that of the
$SL(2)/SL(2)$ topological fied theory,
which has been formulated in \cite{AGS,HY2}. Naturally we might expect the
similarity
between $W_{N}$ strings and
$SL(N)/SL(N)$ topological field theory, as well
as between $N$-extended superconformal and
$OSP(N, 2)/OSP(N, 2)$ topological
field theory.

Our paper is organized as follows.
Section \ref{sec-Gauged} is about gauged  supergroup valued WZNW field theory.
In section \ref{sec-Fe}, it is shown how
$OSP(1, 2)/OSP(1, 2)$ theory is related to the
non-critical fermionic strings.
The relation between different modings of the superalgebra and
their counterparts, the NS and R
superconformal algebra is discussed in detail in section \ref{sec-RNS}.
Finally, in section \ref{sec-The BR}, we formulate the BRST states.
In conclusion, we speculate that the BRST semi-infinite
cohomology of the non-critical fermionic strings can be obtained from
that of $\OSP/\OSP$ by
Hamiltonian reduction approach.

\section{Gauged Supergroup Valued WZNW Model}
\label{sec-Gauged}
The supergroup valued WZNW action can be written as\cite{W1}
\beq
L_k&=&-k/16 \pi\int d^{2}x str \{g^{-1}\partial_{\mu}g g^{-1}\partial_{\nu}g
h^{\mu\nu}\}+\no    \\
  & &+k/24 \pi \int d^{3}x str\{g^{-1}\partial_{\mu}gg^{-1}
\partial_{\nu}gg^{-1}\partial_{\p}g \}\epsilon^{\mu\nu\p},
\label{2.30} \eeq
where $g(z, \bar{z})=e^{i\epsilon^{\af}(z, \bar{z})\tau^{\af}}$ is an
element of a finite dimensional Lie supergroup $G$. $\tau^\af$ is
the generator of the corresponding Lie superalgebra $\cal G$\cite{K0,K2,FY1},
which consists of the even part ${\cal G}_{\bar{0}}$, and the odd part
${\cal G}_{\bar{1}}$. ${\cal G}_{\bar{0}}$ is by itself a
Lie algebra. We shall restrict our discussion to the case that
$\cal G$ is semi-simple
( for definition, see for example ref.\cite{PR,FY1}).
The (anti-)commutators satisfies
\beq
[\tau^\af,\tau^\bt]=(-1)^{d(\af)d(\bt)+1}[\tau^\bt,\tau^\af]=
f^{\af\bt}_{\ \ \ \r}\ \tau^\r,
\eeq
where $f^{\af\bt}_{\ \ \ \r}$ is the structure constant of $\cal G$;
 $d(\af)=0~(~1$ resp.), when
$\tau^{\af}\in {\cal G}_{\bar{0}}~(~{\cal G}_{\bar{1}}$, resp.).
 What needed to point out is
that $\epsilon^{\af}$ is a complex   (Grassmannian resp.) number,
when $d(\af)=0~( ~1$, resp.), so that the group elements commute
cyclicly among themselves
inside the supertrace (see ref.\cite{K0,FY1}),
\beq
str{gh}=str{hg},~~~~g,~h \in G.
\eeq
The left and right conserved currents are defined as
\beq
J=-k/2 \df g g^{-1};~~~~~
\bar {J}=-k/2 g^{-1}\bar{\df}g,
\eeq
which satisfy, through the equation of motion,
\beq
\bar{\df}J=\df \bar{J}=0
.\eeq
Define
\beq
J^{\bt}=str\{\tau^{\bt}J \},\ \ \ \ \
\bar{J}^{\bt}=-str\{\tau^{\bt}\bar{J}\},
\eeq
where the difference between the definitions of $J^\af$ and of $\bar{J}^\af$
by a minus sign ``$-$" comes from the classical Poisson brackets (see
ref.\cite{II}). This also can be seen from the time reversal symmetry of
WZNW theory, $z \leftrightarrow \bar{z},\ g\leftrightarrow g^{-1},
\ J\leftrightarrow -\bar{J}$.
Then we get \cite{W1}
\beq
J^\af(z_1)J^\bt(z_2)&=&\frac{f^{\af\bt}_{\ \ \ \r}J^\r(z_2)}{z_{12}}+
\frac{k\,str\{\tau^\af\tau^\bt\}}{z_{12}^2},\no
\eeq
\beq
{[}J^\af_n,J^\bt_m]&=&f^{\af\bt}_{\ \ \ \r} J^\r_{n+m}+kn\,str\{\tau^\af
\tau^\bt\}\delta_{n+m,0}.
\label{s204}
\eeq
The supertrace is normalized in such a way that when $k$ is a positive
integer, the current algebra eq.(\ref{s204})
has integral representation \cite{GW,K2}. For $G=SU(N),~\OSP$ ( for definition,
see ref.\cite{PR,SNR,FY1}), it is the ordinary (super-)trace in the fundamental
representation.
The same relation holds for the antiholomorphic part.

Using the Polyakov-Wiegmann formula \cite{PW1,PW2},
\beq
L_k(gh)=L_k(g)+L_k(h)-k/2\pi \int d^{2}z\, str(g^{-1}\df_{-}g\df _{+}hh^{-1}),
\eeq
the gauged WZNW action \cite{KS,GK} can be written as
\beq
L_k(g, A, \bar{A})&=&L(g)+k/2\pi \int d^{2}z\,
str(\bar {A} \df_{+}g g^{-1}-g^{-1}\df_{-}gA+\bar{A}gAg^{-1}-\bar{A}A)
   \no\\
&=&L_k(h^{-1}g\td{h})-L_k(h^{-1}\td{h})
,\eeq
where
\beq
A=\df_{+}\td{h}\td{h}^{-1}, &  \no       \\
\bar{A}=\df_{-}hh^{-1}, &\td{h}, h \in H.       \label {3. 1}
\eeq
Here $A, \bar{A}$ are gauge fields, taking values in a subalgebra ${\cal H}$
of ${\cal G}$.
$L(g, A, \bar{A})$ is invariant under the gauge transformation
\beqn\begin{array}{lc}
A\rightarrow\lambda A\lambda^{-1}+\lambda\df\lambda^{-1}, &
\bar{A}\rightarrow\lambda\bar{A}\lambda^{-1}+\lambda\bar{\df}\lambda^{-1}. \\
g\rightarrow\lambda g\lambda^{-1}, &\lambda \in H,
\end{array}\eeqn
Following Polyakov and Wiegmann \cite{PW2}, changing the variables from
$A, \bar{A}$ to $\td{h}, h$, we arrive at the partition function
\beq
Z=\int [dg\,dh\,d\td {h}\,db\,dc]e^{-L_k(h^{-1}g\td{h})+
L_{k+2\td{h}_{H}}(h^{-1}\td{h})-L_{gh}(b, c)}\label{0012}
\eeq
 Fixing the gauge at $h={\bf 1}$, we have
 \beq
Z=\int [dg\,d\td{h}\,db\,dc]e^{-L_k(g\td{h})+
L_{k+2\td{h}_{H}}(\td{h})-L_{gh}(b, c)}\label{0013}
\eeq
where $\td{h}_{H}$ is the dual Coexter number of $H$, which is given in
eq.(\ref{060019}) (We assume the length of the longest root of
$\cal H$ equals that  of $\cal G$), and
\beq
L_{gh}=str(:b\bar{\df}c-\bar{b}\df\bar{c}:),
\eeq
is the action for the ($b$, $c$) ghosts of spin 1 and 0 resp.
When $d(\af)=0~(\ 1, \mbox{resp}.\ )$, ($b_{\af}$, $c^{\af}$) are fermionic (
bosonic, resp.\ ) spin (1, 0) ghosts, and satisfy the same boundary conditions
as the currents
$J^\af$.
\beq
&b=\tau^{\af}b_{\af},~~~~~
c=\tau^{\af}c^{\bt}h_{\af\bt}, \no\\
&\langle b_{\af}(z_{1})c^{\bt}(z_{2})\rangle=
\frac{\delta^{\ \bt}_{\af}}{z_{12}},
\eeq
where $h_{\af\bt}$ is the inverse metric of $h^{\af\bt}$\cite{SNR,FY1}
\beq
h^{\af\bt}=f^{\af\r}_{\ \ \ \rho}f^{\bt\rho}_{~~~\r}(-1)^{d(\rho)}.
\eeq
Sometimes we use notations
\beqn
c_\af=
h_{\af\bt}c^{\bt},~~~~~
b^{\af}=b_{\bt}
h^{\bt\af}.
\eeqn
Of particular interest in our paper is the case
$H=G$, which leads to the $G/G$ model. In that case there are
three sectors of affine Kac-Moody superalgebra of different levels:
\beqn \begin{array}{ll}
J^{\af}(z) &\mbox {with level  }k, \\
\td{J}^{\af}(z)&\mbox {with level  }-k-2\td{h}_{\cal G}, \\
J_{gh}^{\af}(z)&\mbox {with level  }2\td{h}_{\cal G}, \end{array}
\eeqn
where
\beq
&&J_{gh}^{\af}=f^{\af\bt}_{\ \ \ \r}:c^{\r}b_{\bt}:.
\eeq
 For a Lie (super-)algebra, $\td{h}_{\cal G}$
 is the dual Coexter number, and for ${\cal G}=OSP(1,2),~\td{h}_{\cal G}=3/2$.
If $\cal G$ is a finite dimensional
contragradient Lie superalgebra associated with a
Cartan matrix\cite{K2,FY1}, it can be
given by
\beq
&&\td{h}_{\cal G}=\frac{\sum_{\af\in\Delta_{\bar{0}}}|\af|^2-
\sum_{\af\in\Delta_{\bar{1}}}|\af|^2}{r_{\cal G}r_l^2}, \label{060019}
\eeq
where $\Delta_{\bar{\imath}}$ is the set of roots corresponding
to ${\cal G}_{\bar{\imath}}$\cite{K2,FY1}, and $r_l$ is length of the highest
root, which is always in $\Delta_{\bar{0}}$;
$r_{\cal G}$ is the rank of $\cal G$.

Again there remains a twisted $N=2$ superconformal symmetry in this model,
which is the generalized form of that in ref.\cite{HY1}.
\beq
G=(J^\af+\td{J}^\af+\frac{1}{2}J^\af_{gh})c^\bt h_{\af\bt},\ \ \ \ \ \ \
\bar{G}=\frac{\td{h}_{\cal G}(J^\af-\td{J}^\af)b_\af (-1)^{d(\af)}}
{k+\td{h}_{\cal G}},
\no\eeq
\beq
T=\frac{\td{h}_{\cal G}:J^\af J^\bt: h_{\af\bt}}{k+\td{h}_{\cal G}}
-\frac{\td{h}_{\cal G}
:\td{J}^\af \td{J}^\bt: h_{\af\bt}}{k+\td{h}_{\cal G}}+
:\df c^\af b_\af:,
\eeq
\beq
J^{u(1)}=c^\af b_\af
.\no\eeq
The central charges for the stress-energy tensors
are\cite{FY1}
\beq
c=\frac{k\,sdim({\cal G})}{k+\td{h}_{\cal G}},~~~~
\td{c}=\frac{(k+2\td{h}_{\cal G})\,sdim({\cal G})}{k+\td{h}_{\cal G}},~~~~
c_{gh}=-2sdim({\cal G}),
\eeq
where $sdim({\cal G})=dim({\cal G}_{\bar{0}})-dim({\cal G}_{\bar{1}})$.
It is easy to see that the total central charge vanishes.
 The OPEs between the supersymmetry generators are
\beq
G(z_1)G(z_2)=0,\ \bar{G}(z_1)\bar{G}(z_2)=0,
\no\eeq
\beq
G(z_1)\bar{G}(z_2)=
\frac{T^{tot}(z_2)}{z_{12}}+\frac{J^{u(1)}}{z_{12}^2}+
\frac{sdim({\cal G})}{z_{12}^3},
\eeq
while the other super-commutators
are the same as in $N=2$ superconformal field theory.
The main purpose of the present paper is to calculate the physical state space
of this model. As in the standard way, we can reach that by the BRST
approach. Using the above formula it is easy to prove that
\beqn
Q_{BRST}=\oint G   \label{3.100}
\eeqn
satisfies $Q_{BRST}^2=0$. And $T(z),\ J^{tot,\af}$ are total
$Q_{BRST}$ (anti)commutators,
\beqn
T(z)=\{Q_{BRST},G(z)\},\ J^{tot,\af}(z)=
\{Q_{BRST},b^\af (z)\},
\eeqn
We will compute the semi-infinite cohomology in the special case of
${\cal G}=OSP(1,2)$ in section 5.

\section{Noncritical Fermionic Strings and $\OSP/\OSP$ FT}
\label{sec-Fe}
The equivalence between $2$D gravity and $SL(2)/SL(2)$
topological field theory has been worked out by the authors of
ref.\cite{HY1,HY2,AGS}. There is a one to one correspondence between the
physical states
in the two theories when restricting the matter sector
to be in the minimal series.

The relation between 2D supergravity and $\os$ current algebra has been
studied early in ref.\cite{PZ} (see also \cite{BO1}). It is shown that
the correlation functions of super-zweibeins possess an $\os$ symmetry. In
fact the discussion in ref.\cite{HY1} on the equivalence of string theory
and $SL(2)/SL(2)$ field theory can be taken over to
the super case. The super conformal action can be obtained from the
$\OSP$ WZNW theory by imposing constraints on the currents\cite{BO1}.

\subsection{Non-critical Fermionic String}
The action for the fermionic string is\cite{P2,P3}
\beq
S=\frac{1}{2}\int [\sqrt{h}h^{\mu\nu}\partial_\mu X\cdot\partial_\nu X
+ \bar{\psi}\cdot\partial\hspace{-2mm}/\hspace{0.5mm}\psi +\bar{\chi}_\mu
\gamma^\nu\gamma^\mu(\partial_\nu X+\frac{1}{2}\chi_\nu\bar{\psi})\cdot
\psi]d^2\xi
.\eeq
Here, $X^i$ and $\psi^i$ are the world sheet scalar and Majorana fermion resp.,
$\chi_\mu$, the gravitino, is a world sheet
Rarita-Schwinger vector-spinor, $\gamma^\mu$
the Dirac
matrix in two dimensions.
Exploiting the gauge invariance of the action under the reparametrization
and the local supersymmetry transformation, we can reach the following
``superconformal gauge",
\beq
h_{\mu\nu}=e^\varphi\delta_{\mu\nu},\ \ \ \ \ \chi_\mu=\gamma_\mu\chi
.\eeq
In fixing the gauge, there is a Fadeev-Popov determinant involved to match
the gauge volume. This determinant is calculated in ref.\cite{P2,P3} and is
shown to be,
\beq
det(L_F^{-1}L_B)=exp(\frac{10}{32\pi}S(\varphi,\chi))
\int [d\hat{b}d\hat{c}d\hat{\beta}d\hat{\gamma}]
exp\{-S^{[2,-1,3/2,-1/2]}_{gh}(\hat{b},\hat{c},\hat{\beta},\hat{\gamma})\}
.\eeq
Here, ($\hat{b}$,$\hat{c}$) are fermionic ghosts of spin $(2,-1)$ and
($\hat{\beta}$,$\hat{\gamma}$) bosonic ghosts of spin $(3/2, -1/2)$,
\beq
S^{[j,1-j,i,1-i]}_{gh}(b,c,\beta,\gamma)
&=&S_f^{(j,1-j)}(b,c)+S_b^{(i,1-i)}(\beta,\gamma),\no\\
S_f^{(j,1-j)}(b,c)&=&\int (b\bar{\partial}c+h.c.)\no\\
S_b^{(i,1-i)}(\beta,\gamma)&=&\int (\beta\bar{\partial}\gamma +h.c.)
,\eeq
where, $S_f^{(j,1-j)}(b,c)$ ($S_b^{(j,1-j)}(\beta,\gamma)$) is the usual action
for the fermionic $(b, c)$ (bosonic $(\beta, \gamma)$) ghost of spin
$(j, 1-j)$\cite{FMS}.

The form of the super-Liouville action $S(\varphi,\chi)$,
which is the supersymmetrized form of the Liouville action, is
completely determined by the trace anomaly,
\beq
S(\varphi,\chi)=
\int \frac{1}{2}(\partial\varphi)^2+\frac{1}{2}\chi
\partial\hspace{-2mm}/\hspace{0.2mm}\chi
+\frac{1}{2}(\bar{\chi}\gamma_5\chi)e^{\varphi/2}
+\mu^2e^\varphi
.\eeq
In terms of the super-field,
\beq
\phi=\varphi+\theta\bar{\chi}+\bar{\theta}\chi+\theta\bar{\theta}f
,\eeq
we shall consider the following manifestly supersymmetrically invariant action,
\beq
S(\phi)=\int d^2\xi d\theta d\bar{\theta}(\frac{1}{2}D\phi \bar{D}\phi
+2\mu e^{\phi/2})
,\label{0031}\eeq
where, $D=\partial_\theta+\theta\partial_z$,
$\bar{D}=\partial_{\bar{\theta}}+\bar{\theta}\partial_{\bar{z}}$.
Integrating out the odd coordinate $\theta$, $\bar{\theta}$, and
substituting in eq.(\ref{0031}) the minimal value of the field $f$, we recover
the action $S(\varphi,\chi)$ in terms of the component
fields $\phi$ and $\chi$.
The equation of motion for $S(\phi)$ reads,
\beq
D\bar{D}\phi+\mu e^{\phi/2}=0
.\label{0032}\eeq

\subsection{The Hamiltonian Reduction}

Classically, eq.(\ref{0032}) is the constrained form of the equation of
motion for the $\OSP$ WZNW theory
\cite{BO1,II,I} (See eqs.(\ref{006058}) for the super-commutators of $\os$).
The constraints are
\beq
J^+=-\bar{J}^-=1,\ \ \ \ j^+=0,\ \ \ \ \bar{j}^-=0.\label{0001}
\eeq
The constraints eq.(\ref{0001}) are of second class. To make a consistent
quantum field theory, we shall adopt the method used by the
authors in ref.\cite{BO1}.
We first consider the Hamiltonian reduction of the $\os$ superalgebra.
Upon the constraints
\beq
J^+=1,~j^+_>=0,\label{616}
\eeq
an irreducible representation of $\os$ reduces to that
of super-Virasoro algebra. By introducing a free Majorana fermion
$\psi$, the constraints eq.(\ref{616}) are equivalent to the following
constraints on the enlarged space $V_{\os}\otimes V_{\psi}$,
\beq
J^+=-\bar{J}=1,~~~~~j^+=\sqrt{2}\psi, ~~~~~\bar{j}^-=\sqrt{2}\bar{\psi}.
\label{0000}
\eeq
  To make the constraints eq.(\ref{0000}) sensible, i.e.
the currents $J^+,~j^+$ have the
correct conformal dimensions, the stress-energy tensor is modified,
\beq
T(z)\rar T_{impr}(z)=T(z)+\df J^3(z).
\eeq
The reduced space can be got in the standard BRST formalism, in which
one should introduce fermionic ghosts $(b,~c)$ and bosonic ghosts $(\bt,~
\r)$ with conformal isospin $(1,~0)$ and $(1/2,~1/2)$ resp..
The BRST operator is
\beq
Q_{BRST}=\oint dz(b(J^+-1)+\r (j^+-\sqrt{2}\psi)-c\r^2),\label{s502}
\eeq
where the ghosts $(\bt,~\r)$ as well as Majorana
fermion $\psi$ satisfies the same boundary condition as that of $j^\pm(z)$
on the complex plane.
The total stress-energy tensor is
\beq
T^{tot}=T_{impr}-\psi\df \psi/2-c\df b+(\bt\df\r-\df\bt\r)/2,\label{s503}
\eeq
which has central charge
\beq
c^{tot}&=&c^{\os}_{impr}+c^{bc}+c^{\bt\r}+c^{\psi}\no\\
       &=&\frac{k}{k+3/2}-6k-2-1+1/2\no\\
       &=&\frac{3}{2}(1-\frac{8(k+1)^2}{2k+3}).\label{s504}
\eeq

\subsection{Constrained WZNW Theory}
The quantum Hamiltonian reduction considered in the last subsection is
equivalent to the following constrained WZNW theory,
\beq
S(g,A,\psi)&=&S_{WZNW}(g)-\frac{1}{\pi}\int d^2z[\bar{A}_{+}(J^+-1)
+\bar{A}_{\frac{1}{2}}(j^+-\sqrt{2}\psi)+A_{-}(\bar{J}^-+1)
\no\\
&&+A_{-\frac{1}{2}}(\bar{j}^--\sqrt{2}\bar{\psi})
+ str(\bar{A}gAg^{-1})
 +\psi\bar{\partial}\psi/2+
\bar{\psi}\partial\bar{\psi}/2]
,\eeq
where the gauge fields $A, \ \bar{A}$, which are effectively the Lagrange
multipliers, take values in the Borel subalgebra,
\beq
A&=&A_{-\frac{1}{2}}\tau^{-\frac{1}{2}}+A_{-}\tau^{-},\no\\
\bar{A}&=&\bar{A}_{\frac{1}{2}}\tau^{\frac{1}{2}}+\bar{A}_{+}\tau^{+}
.\eeq
$S(g,A,\psi)$ is invariant under the following gauge transformations,
\beq
g&\rar & \lambda g\bar{\lambda}^{-1},\no\\
\bar{A}&\rar&\lambda\bar{A}\lambda^{-1}+
\bar{\partial}\lambda \lambda^{-1}\no\\
A&\rar&\bar{\lambda}A\bar{\lambda}^{-1}+
\partial\bar{\lambda}\bar{\lambda}^{-1}\no\\
\psi&\rar&\psi+\sqrt{2}\epsilon_{\frac{1}{2}}\no\\
\bar{\psi}&\rar&\bar{\psi}+\sqrt{2}\bar{\epsilon}_{-\frac{1}{2}}\no\\
\lambda&=&exp\{\epsilon_{\frac{1}{2}}\tau^{\frac{1}{2}}
+\epsilon_{+}\tau^{+}\}\no\\
\bar{\lambda}&=&exp\{\bar{\epsilon}_{-\frac{1}{2}
}\tau^{-\frac{1}{2}}
+\bar{\epsilon}_{-}\tau^{-}\}
.\eeq
Because of the gauge invariance of the action $S(g, A, \psi)$, we have to
fix the gauge in order that the path integral make sense. For convenience, we
choose our gauge condition to be,
\beq
\bar{A}=A=0
\eeq
The change of measure $[dA d\bar{A}]$ is compensated by the introduction of the
Fadeev-Popov ghost, $(b, c)$ being fermionic of spin (1, 0) and
($\beta$, $\gamma$) being bosonic of spin (1/2, 1/2).
So the gauge fixed path integral for the quantized super-Liouville theory
is
\beq
\int [dg][dbdc][d\beta d\gamma][d\psi]exp\{-S^{twisted}_{WZNW}(g)
-S^{(1,0,1/2,1/2)}_{gh}(b,c,\beta,\gamma)-S(\psi)\}
,\eeq
where, the superscript ``twisted" means that the energy momentum tensor
for the constrained WZNW model is improved as
\beq
T(z)&=&T^{Sugawara}(z)+\partial J^3(z),\no\\
\bar{T}(\bar{z})&=&\bar{T}^{Sugawara}(\bar{z})+\bar{\partial}\bar{J}^3(z).
\eeq
Combining the superconformal matter, super-Liouville and
super-reparametrization ghost together, we arrive at the following path
integral formalism of the non-critical string,
\beq
Z=
&&\int [dg][dbdc][d\beta d\gamma][d\psi]e^{-kS^{twisted}_{WZNW}(g)
-S^{(1,0,1/2,1/2)}_{gh}(b,c,\beta,\gamma)-S(\psi)}\no\\
&&\int [d\tilde{g}][d\tilde{b}d\tilde{c}]
[d\tilde{\beta} d\tilde{\gamma}][d\td{\psi}]
e^{-\tilde{k}S^{twisted}_{WZNW}(\tilde{g})
-S^{(1,0,1/2,1/2)}_{gh}(\tilde{b},\tilde{c},\tilde{\beta},\tilde{\gamma})
-S(\tilde{\psi})}\no\\
&&\int [d\hat{b}d\hat{c}][d\hat{\beta} d\hat{\gamma}]
e^{-S^{(2,-1,3/2,-1/2)}_{gh}(\hat{b},\hat{c},\hat{\beta},\hat{\gamma})}
.\label{0040}\eeq
The total conformal anomaly of the theory should vanish,
\beq
c^{tot}=\frac{k}{k+3/2}-6k-3+1/2
+\frac{\tilde{k}}{\tilde{k}+3/2}-6\tilde{k}-3+1/2
-15=0
,\eeq
which leads to the consistency condition (see ref.\cite{HY1} for a discussion),
\beq
\tilde{k}=-k-3
\eeq
It is worth mentioning that we can reorganize the various ghosts appearing
in the path integral, eq.(\ref{0040}), into the $\os$ multiplets and singlets.
Recall in eq.(16) a level $k=3$ $\os$ Kac-moody current algebra is defined
in terms of the fermionic ghost ($b_a$, $c^a$) of spin (1, 0) and the bosonic
ghost ($\beta_\alpha$, $\gamma^\alpha$) of spin (1, 0), such that we have
the following $sl(2)$ isospin assignment for the ghosts $b_a$, $c^a$,
$\beta_\alpha$, $\gamma^\alpha$,
\beq
\begin{array}{cccccc}
ghosts:&b_-, c^+&b_3, c^3&b_+, c^-&\beta_{-1/2}, \gamma^{1/2}
&\beta_{+1/2}, \gamma^{-1/2}\\
J^3: & 1&0&-1&1/2&-1/2
\end{array}
\no
\eeq
Let us see what happens if the ghost Kac-Moody algebra is also twisted by
$J^{gh,3}$, i.e. the energy momentum tensor for the ghost is improved as
follows,
\beq
T^{gh}(z)&=&\partial c^ab_a+\partial\gamma^\alpha\beta_\alpha+\partial
J^{3,gh}\no\\
\bar{T}^{gh}(\bar{z})&=&\bar{\partial}
\bar{c}^a\bar{b}_a+\bar{\partial}\bar{\gamma}^\alpha\bar{\beta}_\alpha
-\bar{\partial}
\bar{J}^{3,gh}
\eeq
In such a case, the conformal spins of the ghosts will get modified as follows,
\beq
\begin{array}{cccccc}
ghosts:&(b_-, c^-)&(b_3, c^3)&(b_+, c^+)&(\beta_{-\frac{1}{2}},
\gamma^{-\frac{1}{2}})&(\beta_{+\frac{1}{2}}, \gamma^{+\frac{1}{2}})\\
J^3:&(1,-1)&(0,0)&(-1,1)&(1/2,-1/2)&(-1/2,1/2)\\
\mbox{spins before twist }\Delta:&
(1,0)&(1,0)&(1,0)&(1,0)&(1,0)\\
\mbox{spins after twist }\Delta-J^3:&(0,1)&(1,0)&(2,-1)&(1/2,1/2)&(3/2,-1/2)
\end{array}
\no\eeq
After the twist the action for the $\os$ ghosts can be written,
\beq
S^{twited}_{gh}(b_\alpha, c^\alpha)&=&
S^{(2,-1,3/2,-1/2)}_{gh}(b_+,c^+,\beta_{1/2},\gamma^{1/2})+
S^{(1,0,1/2,1/2)}_{gh}(b_3,c^3,\beta_{1/2},\gamma_{-1/2}) \label{06048}\\
&&+S^{(1,0)}_f(c^-,b_-).\no
\eeq
Comparing eq.(\ref{06048}) and eq.(\ref{0013}), we arrive at the following
conclusion,
\beq
\langle|\rangle_{string}&=&\langle|\rangle_{OSP(1,2)/OSP(1,2)}^{twisted}
\int[d\psi^+ d\psi^-][d\beta d\gamma]
e^{S^{(1/2,1/2,1/2,1/2)}_{gh}(\psi^+,\psi^-,\beta,\gamma)}\no\\
\psi^+&=&\psi+i\tilde{\psi}\ \ \ \ \ \psi^-=\psi-i\tilde{\psi}
,\eeq
which is the same as to say that the noncritical fermionic string is equivalent
to the twisted $OSP(1,2)/OSP(1,2)$ gauged WZNW model tensoring a topological
field theory of the spin-(1/2) ($\psi^+,\psi^-,\beta,\gamma$) system.

It is well known that for the gauge fixed action there exists a BRST symmetry.
In our case, we come across a 2d topological conformal field theory, which
means\cite{DVV} that there is a twisted N=2 superconformal algebra. The BRST
charge is just one of the N=2 supersymmetry charge. The N=2 superconformal
algebra for the $OSP(1,2)/OSP(1,2) \otimes (\psi^+,\psi^-,\beta,\gamma)$
theory is constructed as follows,
\beq
&&\begin{array}{ll}
G=(J^\af+\td{J}^\af+\frac{1}{2}J^\af_{gh})c^\bt h_{\af\bt}+\psi^-\beta,&
\bar{G}=\frac{\td{h}_{\cal G}(J^\af-\td{J}^\af)b_\af (-1)^{d(\af)}}
{k+\td{h}_{\cal G}}+\frac{1}{2}(\psi^+\partial\gamma-\partial\psi^+\gamma),\\
T=\frac{\td{h}_{\cal G}:J^\af J^\bt: h_{\af\bt}}{k+\td{h}_{\cal G}},&
\td{T}=-\frac{\td{h}_{\cal G}
:\td{J}^\af \td{J}^\bt: h_{\af\bt}}{k+\td{h}_{\cal G}},\\
T^{gh}=\df c^\af b_\af,&J^{u(1)}=c^\af b_\af +\psi^-\psi^+/2-\beta\gamma/2,\\
\end{array}\no\\
&&T^{(\psi^+,\psi^-,\beta,\gamma)}=
\frac{1}{2}(\partial\psi^+\psi^--\psi^+\partial\psi^-)
+\frac{1}{2}(\beta\partial\gamma-\partial\beta\gamma),\\
&&T^{tot}=T+\td{T}+T^{gh}+T^{(\psi^+,\psi^-,\beta,\gamma)}.\no
\eeq
The central charge for the total stress-energy tensor still vanishes.

Using the above formula it is easy to prove that
\beqn
Q_{BRST}=\oint G   \label{3.100'}
\eeqn
satisfies $Q_{BRST}^2=0$. And $T(z),\ J^{tot,\af}$ are total
$Q_{BRST}$ (anti)commutators,
\beqn
T(z)=\{Q_{BRST},G(z)\},\ J^{tot,\af}(z)=
\{Q_{BRST},b^\af(z)\}.
\eeqn
The total BRST charge can be rewritten as
\beq
Q_{BRST}=Q_{BRST}^{\os}+Q_2,
\eeq
where $Q_{BRST}^{\os}$ is the BRST charge for the gauged $\OSP/\OSP$ current as
given in eq.({\ref{3.100}),
and $Q_2$ is the BRST charge for the $(\psi^+,\psi^-,\bt,\r)$ system,
\beq
Q_2=\oint \psi^-\bt \eeq
 The BRST state space of $Q_2$
consists of only one cohomology class which is represented by
the vacuum of the $(\psi^+,\psi^-,\bt,\r)$ system. From
\beq
T^{\psi^+,\psi^-,\bt,\r}=\{Q_2,\ (\psi^+\df\r-\df\psi^+\r)/2\},
\eeq
we come to the fact that the nontrivial $Q_2$ state must be of
zero mode excitation. If the fields $\psi^+,\ \psi^-,\ \bt,\ \r$ are periodic
on the plane, there is no zero mode generators.
However if they are antiperiodic, from
\beq
n_{\psi^-_0}+n_{\r_0}=\psi^-_0\psi^+_0+\r_0\bt_0=\{Q_2,\ \psi^+_0\r_0\},
\eeq
it can be seen that nontrivial  $Q_2$ states must satisfy $n_{\psi^-_0}
+n_{\r_0}=0$, which leads to the choice of two vacuum states which are
not connected by a finite number of zero mode actions.
In both cases the only BRST state is the vacuum state. Now
it is obvious that the total BRST state  space is the direct product of
that of the  $Q_{BRST}^{\os}$ and that of $Q_2$, i.e.
\beq
H(V^{\os}\otimes (\psi^+,\psi^-,\bt,\r),\ Q_{BRST})\cong H(V^{\os},\
Q_{BRST}^{\os})\otimes |vac\rangle_{\psi^+,\psi^-,\bt,\r}.
\eeq
We will compute the semi-infinite cohomology $H(V^{\os},\
Q_{BRST}^{\os})$ in section
\ref{sec-The BR}, after reviewing some results obtained in our
previous paper\cite{FY1} about $\os$.

\section{R type and NS type $\os$}
\label{sec-RNS}
In (extended) superconformal field theories, there are different modings
for the supersymmetry generators, the so called Neveu-Schwarz
(half integral moding) or Ramond (integral moding) sectors. Since the
(extended) superconformal algebra can be considered as the reduced theory
of the corresponding supergroup valued WZNW model, it deserves a special
attention to consider how the different sectors in the formal case are
related to the different types of the superalgebra in the later one.

In this section, we shall restrict ourselves to the case of $N=1$
superconformal algebra and the $\os$ Kac-Moody algebra, although our
consideration is generalizable to the other cases.

The superalgebra $\os$ consists of the $\widehat{SL(2)}$ and the fermionic
part $\{j_r^\pm\}$\cite{FY1}. The nonvanishing (anti-)commutators are
\beq
&\begin{array}{ll}
 \{ j^{+}_r, j^{-}_s \}=2J^{3}_{r+s}+2rk\delta_{r+s,0};
& \{j^{\pm}_r, j^{\pm}_s\}=\pm2J^{\pm}_{r+s};  \\
{[}J^{3}_n, j^{\pm}_r]=\pm \frac{1}{2}j^{\pm}_{n+r};
 &[J^{\pm}_n, j^{\mp}_r]=-j^{\pm}_{n+r};  \\
 {[}J^{+}_n, J^{-}_m]=2J^{3}_{n+m}+nk\delta_{n+m,0};
 &[J^{3}_n, J^{\pm}_m]=\pm j^{\pm}_{n+m}, \end{array}\label{006058}\\
                 &{[}d,J_n^\af]=-nJ_n^\af. \no
\eeq
The stress-energy tensor by Sugawara construction is
\beq
T(z)=\frac{:J^3J^3+J^+J^-/2+J^-J^+/2-j^+j^-/4+j^-j^+/4:(z)}{k+3/2},
\label{s400}
\eeq
with central charge
\beq
c=\frac{k}{k+3/2}.
\eeq
 Two types of $\os$ algebra, which we call Ramond type and
Neveu-Schwarz type $\os$ algebra resp (or in terms of ref.\cite{BO1},
the untwisted and twisted $\os$
current algebra, resp.), were studied in ref.\cite{FY1}.
The fermionic generators of $\os$, i.e.\
$j^{\pm}_r$ are of integer (half integer) modes for the R (NS) type
$\os$.
It was shown there that, although the NS and R super Virasoro algebra are
genuinely different, however,
the two types of $\os$
are essentially identical via an isomorphic map, so are the representations
of the two types of $\os$ superalgebra. This isomorphism is the
main part of this
section. The isomorphism on the superalgebras are
\beq
\begin{array}{cc}
j_n^{\pm}\rar \pm j^{\mp}_{n\pm 1/2},&J_n^{\pm}\rar -J_{n\pm 1}^{\mp},\\
J_n^3\rar-J_n^3+k/2\delta_{n,0},&k\rar k,\\
d\rar d+J_0^3.\end{array} \label{s401}
\eeq
For convenience we add a superscript R or NS on the generators,  the
modules, and the corresponding weights when we are concerned about
the R or NS type $\os$. From the above isomorphism eq.(\ref{s401}),
we have
\beq
j^{R}=k/2-j^{NS},\label{s505}
\eeq
which gives the relation between the isospins of a certain state
in the representations of R and  NS type $\os$.

However, the R and NS type $\os$ current algebra, which correspond to the
periodic and antiperiodic boundary conditions for the fermionic currents resp.,
have different behaviour on
the complex plane.
Physically the Virasoro algebra eq.(\ref{s400}) by Sugawara construction are
different,
\beq
L_n^R=L_n^{NS}-J_n^{3,NS}+k/4\delta_{n,0}.\label{s506}
\eeq
When the Hilbert
 space is concerned, a representation of the $\os$ algebra corresponds
to two family of states, one in R sector, another in NS sector. At least at
the level of current algebra, we can say that the conformal blocks are
degenerated.

On the torus, there are two homotopically nontrivial cycles, the $a$ cycle
and $b$ cycle resp.. Correspondingly, there are four different boundary
conditions for the fermionic generators. The boundary conditions along
the $\sigma$ direction are specified by the moding of the generators,
while along $\tau$ direction, the boundary condition is reflected by
defining the characters or the super-characters. For more detailed discussion,
see ref.\cite{FY1}.

The structures of the Verma modules and Wakimoto modules over $\os$
are very much  similar
to those over Virasoro and over $\widehat{SL(2)}$. The modules can also
be classified into cases I, II, III.
Of special interest is the admissible representations (in case III$_-$), where
\beq
 &&2k+3=\td{q}/q,~~q, ~\td{q} \in {\bf N},~~q+\td{q}\in\mbox{even},~~
 gcd(q,~\frac{q+\td{q}}{2})=1;\no\\
&&4j_{m,s}^R+1=m-s\frac{\td{q}}{q},~~m=1,\ldots, \td{q}-1,~~s=0,\ldots,q-1,~~
m+s\in \mbox{odd}.\label{s509}
\eeq
Notice that when expressed in the NS type $\os$, the isospin in
eq.(\ref{s509}) can be rewritten as, by using eq.(\ref{s505}),
\beq
4j^{NS}+1=(\td{q}-1-m)-(q-1-s)\frac{\td{q}}{q}.\label{s424}
\eeq
The corresponding Verma module $M(j_{m,s}^R,~k)$ has singular vector with
isospin $j_{2n\td{q}\pm m, s}$. The Wakimoto module $W(j^R_{m,s},~k)$,
which is realized in free fields \cite{Wo,BO1,FY1},
has singular vectors with isospin $j_{2n\td{q}-m, s}, ~n>0$,
cosingular vectors with $j_{2n\td{q}-m, s}, ~n<0$, and when modulo
the submodules generated by the singular
vectors, there appear singular vectors with $j_{2n\td{q}+m, s}$ in
$W(j^R_{m,s},~k)$.

The free field realization of $\os$\cite{BO1,FY1}
 with the Fock space $F_{j,\af}$  being $W(j,k)$ is
\beq
&&\left\{ \begin{array}{lll}
         J^+&=&-\bt,\\
         j^+&=&\psi^+-\bt\psi,\\
         J^3&=&-\bt\r+i\af_+\df\phi/2-\frac{1}{2}\psi\psi^+-\epsilon/4z,\\
         j^-&=&\r(\psi^+-\bt\psi)+i\af_+\psi\df\phi+(2k+1)\df\psi,\\
         J^-&=&\bt\r^2-i\af_+\r\df\phi+\r\psi\psi^+-
k\df\r+(k+1)\psi\df\psi+\epsilon\r/2z,
 \end{array}\right.\label{s404}\\
&&T=\bt\df\r-\psi^+\df\psi-(\df \phi)^2/2-i\df^2\phi/(2\af_+)-\epsilon/8z^2
\no
\eeq
where $\af^2_+=2k+3$, and $\epsilon=0,~1$ for R type or NS type $\os$ resp.;
$T(z)$ is the stress-energy tensor
by Sugawara construction eq.(\ref{s400}). From the
 involution $\sigma$ of $\os$,
\beq
\begin{array}{cc}
\sigma(J_n^\pm)=-J_n^\mp,&\sigma(j_n^\pm)=\pm j_n^\mp,\\
\sigma(J_n^3)=-J_n^3,&\sigma(k)=k,~\sigma(d)=d,
\end{array}\label{s426}
\eeq
we get another free field realization,
\beq
&&\left\{ \begin{array}{lll}
        \td{J}^+&=&-\td{\bt}\td{\r}^2+i\td{\af}_+\tilde{\r}\df\td{\phi}-\td{\r}
\td{\psi}\td{\psi}^++
\td{k}\df\td{\r}-(\td{k}+1)\td{\psi}\df\td{\psi}-\epsilon\r/2z,\\
         \td{j}^+&=&-\td{\r}(\td{\psi}^+-\td{\bt}\td{\psi})-i\td{\af}_+
\td{\psi}\df\td{\phi}-(2\td{k}+1)\df\td{\psi},\\
         \td{J}^3&=&\td{\bt}\td{\r}-i\td{\af}_+
\df\td{\phi}/2+\frac{1}{2}\td{\psi}\td{\psi}^++\epsilon/4z,\\
         \td{j}^-&=&\td{\psi}^+-\td{\bt}\td{\psi},\\
         \td{J}^-&=&\td{\bt},
 \end{array}\right.\label{s405}\\
&&\td{T}=\td{\bt}\df\td{\r}-\td{\psi^+}\df\td{\psi}-(\df\td{ \phi})^2/2
-i\df^2\td{\phi}/(2\td{\af_+})-\epsilon/8z^2
\no
\eeq
The Fock space $F_{j,\af}$ ($\td{F}_{\td{j},\td{\af}}$) is generated
by negative modes of these fields together with
$\r_0,\ \psi_0$ ($\tilde{\bt}_0,~\td{\psi}_0^+$).
It can be verified that in fact $F_{j,\af}$ and
$\td{F}_{j,\af}$ are dual spaces of each other. The inner product can be
defined as $(|vac\rangle_F,|vac\rangle_{\td{F}})=1$, and satisfies
\beqn
\begin{array}{cc}
(\bt_nu,~ v)=-(u,\td{\bt}_{-n}v),&(\r_nu,~ v)=(u,\td{\r}_{-n}v),\\
(\psi_nu,~ v)=-i(u,\td{\psi}_{-n}v),&(\psi^+_nu,~ v)=
-i(u,\td{\psi}^+_{-n}v),\\
(\phi_nu,~ v)=-(u,\td{\phi}_{-n}v),&\mbox{For~~}u\in F_{j,\af},~v\in
\td{F}_{j,\af}.
\end{array}
\eeqn
The inner product so defined satisfies
\beqn
\begin{array}{cc}
(J^\pm_nu,~ v)=(u,\td{J}^\mp_{-n}v),&(j^\pm_nu,~ v)
=\pm i(u,\td{j}^\mp_{-n}v),\\
(J^3_nu,~ v)=(u,\td{J}^3_{-n}v),&\mbox{For~~}u\in F_{j,\af},~v\in
\td{F}_{j,\af}.
\end{array}
\eeqn
In our following discussion, we use notation $W(j,k)$ to denote $F_{j,\af}$,
and  $W^*(j,k)$ to denote $\td{F}_{j,\af}$.

Here we show that the isomorphism between R and NS type $\os$
can be realized by a map between the free fields. We still add a superscript
R, or NS to the free fields when it is in the  free field realization of the
R, or NS type $\os$.
Let
\beq
\begin{array}{cc}
\psi^{+,R}_{n}\rar\psi^{+,NS}_{n+1/2},&\psi_n^R\rar\psi^{NS}_{n-1/2},\\
\bt_n^R\rar\bt_{n+1}^{NS},&\r_n^R\rar\r_{n-1}^{NS},\\
\phi^R_n\rar\phi_n^{NS}+\af_+/2\delta_{n,0}
\end{array}
\eeq
i.e.
\beq
\begin{array}{cc}
\psi^{+,R}\rar z^{1/2}\psi^{+,NS},&\psi^R\rar z^{-1/2}\psi^{NS},\\
\bt^R\rar z\bt^{NS},&\r^R\rar z^{-1}\r^{NS},\\
i\df\phi^R\rar i\df\phi_n^{NS}+\af_+/2z.
\end{array}
\eeq
It is consistent with the map eq.(\ref{s401}), and the realization of R type
$\os$ in eq.(\ref{s404}) becomes  the dual realization of NS type $\os$
in eq.(\ref{s405}).

Now we consider the ghost spaces. $gh^{R,(0,\frac{1}{2})}$ denotes R type
ghost space with vacuum
$|\rangle _{(0,\frac{1}{2})}^R$ annihilated
by $c_0^+$ and $c_0^{1/2}$; and $gh^{NS,(1,1)}$ denotes NS one with
vacuum $|\rangle_{(1,1)}^{NS}$ annihilated by $c_0^+$. The scripts
$(0,\frac{1}{2})$ and $(1,1)$ correspond to $(J^{u(1)}_0, J^{gh, 3}_0)$
of the ghost vacua.
It can be verified that $gh^{R,(0,\frac{1}{2})}$ and
$gh^{NS,(1,1)}$ are also isomorphic, as can be seen from the following map
\beqn
\begin{array}{cc}
b_-^R\rar zb_+^{NS},&c^{-,R}\rar z^{-1}c^{+,NS},\\
b_{-1/2}^R\rar z^{1/2}b_{1/2}^{NS},&c^{-1/2,R}\rar z^{-1/2}c^{+1/2,NS},\\
b_3^R\rar b_+^{NS},&c^{3,R}\rar c^{3,NS},\\
b_{1/2}^R\rar -z^{-1/2}b_+^{NS},&c^{1/2,R}\rar -z^{1/2}c^{+1/2,NS},\\
b_+^R\rar z^{-1}b_+^{NS},&c^{+,R}\rar zc^{-,NS}.
\end{array}\eeqn
under which the BRST operator $Q_{BRST}^R$ becomes  $-Q_{BRST}^{NS}$.
{}From the above discussion we have the following proposition
\bpp
\beq
\begin{array}{cc}
M^R(j^R,~k)\cong M^{NS}(j^{NS},~k),&
L^R(j^R,~k)\cong L^{NS}(j^{NS},~k),\\
W^R(j^R,~k)\cong W^{*,NS}(j^{NS},~k),&
gh^{R,(0,\frac{1}{2})}  \cong gh^{NS,(1,1)},
\end{array}\label{s433}
\eeq
\epp
where $j^R$ and $j^{NS}$ are related by  eq.(\ref{s505}).

The difference between the two types of $\os$ algebra becomes more obvious
under the Hamiltonian reduction. Naturally, R type ( NS type, resp.) will
result in NS sector ( R sector, resp.) super-Virasoro algebra due to
the fact that they satisfy the same boundary condition on the complex plane.
The twisting term
$\df J^3$ in $T_{impr}$ brings about a shift  of the
conformal weight, as well as the modings of the fermionic generators
by half integer.

An HWS in R type $\os$ with isospin $J^R$ becomes the HWS in
 the representation of super-Virasoro algebra with conformal weight
\beq
h=\frac{j^R(j^R+1/2)}{k+3/2}-j^R.\label{s507}
\eeq
We should be more careful when considering the conformal weight of the reduced
HWS which is originally an HWS of NS type $\os$. Notice that here the
bosonic ghosts $(\bt,~\r)$ and Majorana fermion $\psi$ are all antiperiodic
on the complex plane. Thus the vacuum have conformal weights $-1/8$ for
$(\bt,~\r)$ system and $1/16$ for fermion $\psi$.
Precisely because that the twisting of the Majorana fermion field $\psi$
is genuine, the NS and R super-Virasoro algebras, as the results of the
Hamiltonian reduction of the R type and NS type of the
$\os$ algebra resp., are no longer isomorphic.
The conformal weight of an HWS in NS type $\os$ with isospin $j^{NS}$
(related to the VirasoroB
algebra by Sugawara construction in eq.(\ref{s400}), no $\df J^3$ term) is
\beq
h^{\os,NS}=\frac{j^{NS}(j^{NS}+1)}{k+3/2}-c/8,\label{s5006}
\eeq
where $c$ is the central charge, $c=\frac{k}{k+3/2}$. Eq.(\ref{s5006})
can also be obtained by using eqs.(\ref{s505},\ref{s506},\ref{s507}).
Now we come to the conformal weight of such an HWS in the super-Virasoro
algebra,
\beq
h=\frac{j^{NS}(j^{NS}+1)}{k+3/2}-j^{NS}-\frac{1}{8}\frac{k}{k+3/2}-1/8+1/16.
\label{s508}
\eeq
Combining eq.(\ref{s504}) and eq.(\ref{s509}) we get
the central charge for the super-Virasoro algebra,
\beq
c^{tot}=\frac{3}{2}(1-\frac{2(\td{q}-q)^2}{q\td{q}})=c_{\td{q},q};
\eeq
By eqs.(\ref{s509},\ref{s424},\ref{s507},\ref{s508}), we get the
conformal weights for the HWS's of the admissible representations
under the Hamiltonian reductin,
those in
\beq
h^R&=&\frac{(mq-(s+1)\td{q})^2-(q-\td{q})^2}{8q\td{q}}=\Delta_{m,s+1};\no\\
h^{NS}&=&\frac{(mq-s\td{q})^2-(q-\td{q})^2}{8q\td{q}}+1/16=\Delta_{m,s}.
\eeq
 We see that they coincide
with those of the super-Virasoro algebra in
the ($q,~\td{q}$) minimal model except for those
 $s=q-1$ in the R type and $s=0$ in NS type $\os$.
Moreover, if we set $p=q+1$, we get the unitary part
\beq
c&=&3/2(1-\frac{8}{q(q+2)}),\\
h^R&=&\frac{(mq-(s+1)(q+2))^2-4}{8q(q+2)},\no\\
h^{NS}&=&\frac{(mq-s(q+2))^2-4}{8q(q+2)}+1/16.
\eeq
So it is easily seen that the admissible representation $L_{m,s}^{R,\os}$
reduces to $L^{NS,Vir}(c_{q,\td{q}},\Delta_{m,s+1})$ and $L_{m,s}^{NS,\os}$,
with highest weight $j_{m,s}^{NS}=k/2-j_{m,s}^R$
 to $L^{R,Vir}(c_{q,\td{q}},\Delta_{m,s})$. Notice that
$L(c_{\td{q},q},\Delta_{m,s})\cong L(c_{\td{q},q},\Delta_{\td{q}-m,q-s})$
while their original images are not isomorphic. The Wakimoto module of the
matter sector realized in the free fields, eq.(\ref{s404}), reduces to the Fock
space $F_{\xi,\eta}$ (in terms of ref.\cite{LZ5}) with $\xi=(\af_+^2
-1)/(2\af_+),~\eta=-(4j+1+\epsilon-\af_+^2)/(2\af_+)$. The dual Wakimoto
module of the
Liouville sector in eq.(\ref{s405}) reduces to that with $\xi=(\af_+^2
-1)/(2\af_+),~\eta=(4j+1+\epsilon-\af_+^2)/(2\af_+)$. Again the reduced Fock
space
$F_{\xi,\eta}$ and $F_{\xi,-\eta}$ are dual spaces of each other.

\section{The BRST States}\label{sec-The BR}

The physical state space under constraints can be obtained by
BRST approach. To obtain the physical states in the $\OSP/\OSP$ gauged
WZNW field theory, in this section
we manage to calculate the BRST cohomology with coefficient
in $\os$ modules. To make our discussion selfconsistent, some results
in (co)homology theory are presented along with the sketch of the proofs
(see also appendix for some technical details). Readers not familiar with
our notations are referred to refs.\cite{LZ2,LZ5,BMP3} for more mathematical
background.

The BRST charge is given in
eq.(\ref{3.100}), explicitly
\beq
Q=\sum_n c_{\af,n}(J_{-n}^{\af}+\td{J}_{-n}^{\af})-1/2\sum_{n,m}
f^{\af\bt}_{~~~\r}c_{\af,n}c_{\bt,m}b^{\r}_{-n-m}.
\eeq
We always
assume that the ghost vacuum $|gh\rangle_0$, which has ghost number zero
by our notation, is annihilated by all the positive modes of the
$b_\alpha,\ c^\alpha$ as well as the the zero modes $b_{\af,0}$
unless under special declaration.
The other kinds of ghost vacua, for example, the one used in ref.\cite{LZ2},
which is annihilated by $c_0^{\frac{1}{2}}$ and $c_0^+$ and is denoted by
$|gh\rangle_{(0,\frac{1}{2})}$ in our paper, can not be obtained
from $|gh\rangle_0$ by a finite number of operations of the ghost modes,
due to the fact that $c_0^{1/2}$ is bosonic. However, by fermionizing the
bosonic ghosts as what was done in ref.\cite{FMS}, ghost vacua of different
bosonic ghost numbers are interpolated by a vertex operator..

As in the standard procedure
of formulating the semi-infinite cohomology, we first consider the relative
complex.
\beqn
C^*_{rel}=\{w|w\in C^*,\ J_0^{tot,3}w=L_0w=b_{3,0}w=0\}
\eeqn
Let
\beqn
Q=Mb_{3,0}+c_0^3J_0^{tot,3}+\hat{Q},
\eeqn
\beqn
M=\sum_n(:c_n^+c_{-n}^-+1/4c_n^{\frac{1}{2}}c_{-n}^{-\frac{1}{2}}:),
\eeqn
and $\hat{Q}$ has no term containing $c_0^3,b_{3,0}$.
When restricted to the relative complex, $Q=\hat{Q}$.
We pay our attention mainly on the
following relative cohomology:
\[ (i)\ \h{*}_{rel}(L(j_1,k_1)\otimes M(j_2,k_2));\
 (i\!i)\ \h{*}_{rel}(W(j_1,k_1)\otimes W^*(j_2,k_2));\]
\[(i\!i\!i)\ \h{*}_{rel}(L(j_1,k_1)\otimes W(j_2,k_2));\
(i\!v)\ \h{*}_{rel}(L(j_1,k_1)\otimes L(j_2,k_2));\]
where we have put on the restriction that $k_1+k_2+3/2=0$, which must
be satisfied for the nilpotency of the BRST operator.
We have to consider another set of $\os$-modules, by abuse of language,
which are called ``lowest weight modules" ( LWM ). The vacuum vector of such
a module is annihilated by $j_0^-,\ J_0^-$ and all $J_n^\af$'s with $n>0$.
We also have LW Verma modules, Wakimoto modules and irreducible modules,
which are denoted by $\td{M}(j,k),\ \td{W}(j,k)$ and $\td{L}(j,k)$
respectively.

We adopt the notation $(j_2,k)\rightarrow (j_1,k)$, to denote that there exists
an embedding $M(j_2,k)\rightarrow M(j_1,k)$, which is equivalent to
the fact that there
exists a singular vector with isospin $j_2$ in  $M(j_1,k)$.
If $(j_2,k)\rightarrow (j_1,k)$, define $l((j_1,k),(j_2,k))=d$, where $d$ is
the
maximal number such that there exists $(j_2,k)=(j_1',k)\rar(j_2',k)\rar
\ldots\rar(j_{d+1}',k)=(j_1,k)$, where $j_i'\neq j_j'$; and
$l((j_1,k),(j_1,k))=0$. If $M(j,k)$ is in the case III$_\pm$, define $j^{\pm
i}$
be such that $l((j,k),(j^{\pm i},k))=|i|,~j^i\neq j^{-i}$. (These definitions
are analogous to that in ref.\cite{LZ5}).
Sometimes for a given level we may omit $k$ in the above definition
when making no confusions.

Our results again is similar to that in $2D$ gravity, given in
ref.\cite{LZ3,BMP1}.
It will be helpful to list two theorems in homology theory which are
relevant to our work. One is the
twisted reduction formula (cf. reduction formula in ref.\cite{LZ2}),
which is proved in the appendix, where the readers are referred to find
some notations used in our calculation here. Another is
for the double cohomology (see, for example \cite{HS,BT,BMP3}).
\btm \label{tm05}
[Twisted Reduction formula]
Let ${\cal G}=\os$, $V$ is a $\cal G$-module in the category
$\cal O$, there is a canonical isomorphism
\beqn
\h{n+1}_{rel}(V\otimes M(-\bt -\lambda)\otimes gh, \hat{Q}
) \cong H^n(V\otimes gh_+,Q_+)[\lambda]; \label{061501}
\eeqn
where $\bt=2\rho=\frac{1}{2} J^{3'}_0+3k'.\ (j'(j)=1,\ k'(k)=1)$;
$[\lambda]$ means restriction to the subspace with weight $\lambda$.
\etm

When one factor of the tensor $V_1\otimes V_2$
is an irreducible module, we have to consider the double
cohomology again and again. The following theorem will be powerful
for our calculation \cite{HS,BT,BMP3}.
\btm\label{tm02}
Let $(c^{p,q},d,\partial)$ be a double complex, which satisfies
\[ d:\ c^{p,q}\rightarrow c^{p+1,q},\
\partial:\ c^{p,q}\rightarrow c^{p,q+1},\
 \{d,\partial\}=0, \]
If the two corresponding sequence both collapse at the second term.
then
\beqn
\sum_{p+q=n} H^q(H^p(c^{*,*},d),\partial)\cong
\sum_{p+q=n} H^q(H^p(c^{*,*},\partial),d).
        \label{7.10}
\eeqn
\etm
In fact our complex is rather simple, many times the following corollary
is enough.
\bc \label{c03}
If both the cohomologies corresponding to $d$ and $\partial$ vanishes
for all but one degree, then (\ref{7.10}) holds.
\ec
First we consider the BRST states of the R type $\os$. However the BRST states
of the NS type can be easily obtained by the isomorphism
eqs.(\ref{s401},\ref{s433}) between them.
\btm \label{tm07}
(i) Let $M_j$ not in case $I\!I\!I^0_{\pm}$, then
\beqn
\h{n+1}_{rel}(L(j_1,k_1)\otimes M(j_2,k_2))\cong
\left\{ \begin{array}{ll}\delta_{n,l(j_1,-1/2-j_2)}{\cal C},&
\mbox{if }-\frac{1}{2}-j_2\rightarrow j_1;\\
\emptyset,&\mbox{otherwise}
\end{array}    \right.
\eeqn
$(i\!i)$ for $M_j$ in case $I\!I\!I^0_{\pm}$, then
\beqn
\h{n+1}_{rel}(L(j_1,k_1)\otimes M(j_2,k_2))\cong
\left\{ \begin{array}{ll}\delta_{n,0}{\cal C},&
\mbox{if }j_1+j_2+\frac{1}{2}=0;\\
\delta_{n,1}{\cal C},& \mbox{if } -\frac{1}{2}-j_2\rar j_1,
\mbox{and }l(j_1,-1/2-j_2)=1,\\
\emptyset,&\mbox{otherwise}
       \end{array} \right.
\eeqn
\etm
To prove the theorem we first prove the following lemma
\blm
(i) Let $M(j,k)$ not in case $I\!I\!I^0_{\pm}$, then
\beqn
H^n(L(j,k)\otimes gh_-,Q_-)\cong \oplus_{j'\rightarrow j,l(j,j')=-n}
           {\cal C}v_{j'};
\eeqn
($i\!i$) for $M(j,k)$ in case $I\!I\!I^0_{\pm}$, then
\beqn
H^n(L(j,k)\otimes gh_-,Q_-)\cong \oplus_{j'\rightarrow j,l(j,j')=n}
           {\cal C}v_{j'}(\delta_{n,0}+\delta_{n,-1}),
\eeqn
\elm
where
$v_j$ is a singular vector with isospin $j$ and conformal weight
$\frac{j(j+1/2)}{k+2/3}$.\\
Proof. $L(j,k)$ is a resolution of  Verma modules,
\beqn
\cdots \stackrel{\partial}{\rar} M^{i}\stackrel{\partial}{\rar}
M^{i+1}\stackrel{\partial}{\rar} \cdots \stackrel{\partial}{\rar}
M^{-1} \stackrel{\partial}{\rar} M^0=M(j,k)\rar 0
\eeqn
where $M^{i}$ is a direct sum of Verma modules.
for a Verma module $M(j,k)$, $H^n({\cal G},M(j,k),Q_-)=\delta_{n,0}{\cal
C}v_j$.
We see that the problem in question is actually  a double complex which
satisfies the condition in the corollary. So
\beqn
H^n(L(j,k)\otimes gh_-,Q_-)\cong H^n(H^0(M^*,\partial)\otimes gh_-,Q_-)\\
=H^n(H^0(M^*\otimes gh_-,Q_-),\partial)
\label{7.20}
\eeqn
According to the fact that $Q_-,\df$ do not change the weight of
vectors( i.e., they are weight zero operators), the $\df$ action
in the right hand side (r.h.s.) of eq.(\ref{7.20})
 is trivial. This completes the proof
of the lemma.

Proof of the theorem. By reduction formula
\beqn
\h{n}_{rel}(L(j_1,k_1)\otimes M(j_2,k_2))\cong
H^n(L(j_1,k_1)\otimes gh_+,Q_+)[\lambda],
\eeqn
where $\lambda=(-1/2-j_2)(J^{3}_{0})'+\frac{j_2(j_2+1/2)}{k+3/2}L_0'$.
For the irreducible module $L(j,k)$ satisfies $L(j,k)\cong L(j,k)^*$,
 so by Poincar\'{e} duality theorem
\beqn
H^n(L(j,k)\otimes gh_+,Q_+)=(H^{-n}(L(j,k)\otimes gh_-,Q_-))^*,
\eeqn
the theorem follows.

By theorem \ref{tm07}, we can compute the cohomology
$\h{n}_{rel}(L(j_1,k_1)\otimes L(j_2,k_2))$ easily.
\btm
($i$) For $M(j_1,k_1)$ in the case ($I$),
\beqn
\h{n+1}_{rel}(L(j_1,k_1)\otimes L(j_2,k_2))\cong {\cal C}\delta_{n,0}
\delta_{j_1+j_2+1/2,0}.
\eeqn
($i\!i$) For $M(j_1,k_1)$ in the case ($I\!I$),
\beqn
\h{n+1}_{rel}(L(j_1,k_1)\otimes L(j_2,k_2))\cong
\left\{\begin{array}{ll} {\cal C}\delta_{n,0}, &\mbox{if } j_1+j_2+1/2=0;\\
{\cal C}(\delta_{n,-1}+\delta_{n,1}),&\mbox{if }-1/2-j_2\rar j_1,\\
&\mbox{ and }j_1+j_2+1/2\neq 0;\\
\emptyset,&\mbox{otherwise}. \end{array}\right.
\eeqn
($i\!i\!i$) For $M(j_1,k_1)$ in the case ($I\!I\!I^0_{\pm}$),
 \beqn
\h{n+1}_{rel}(L(j_1,k_1)\otimes L(j_2,k_2))\cong
\left\{\begin{array}{ll} {\cal C}\delta_{n,0}, &\mbox{if }
-j_2-1/2\rar j_1,\\
&\mbox{ and }l(j_1,-j_2-1/2)=0,\mbox{ or }2;\\
{\cal C}(\delta_{n,-1}+\delta_{n,1}),&\mbox{if }-1/2-j_2\rar j_1,\\
&\mbox{ and }l(j_1,-j_2-1/2)=1;\\
\emptyset,&\mbox{otherwise}. \end{array}\right.
\eeqn
($i\!v$) For $M(j_1,k_1)$ in the case ($I\!I\!I_{\pm}$),
\beqn
\h{n+1}_{rel}(L(j_1,k_1)\otimes L(j_2,k_2))\cong
\left\{\begin{array}{l} {\cal C}(1+(-1)^{n+l(j_1,-1/2-j_2)}-
\delta_{|n|,l(j_1,-1/2-j_2)})
, \\ \mbox{     if } -j_2-1/2\rar j_1,\mbox{ and }|n|\leq l(j_1,-1/2-j_2);\\
\emptyset,\mbox{   otherwise}. \end{array}\right. \label{6.90}
\eeqn
\etm
Proof. ($i$) is obvious.\\
($i\!i$) and ($i\!i\!i$).
 If $M(j_1,k_1)$ in the case $(I\!I)$ or $(I\!I\!I^0_\pm)$, the resolution of
$L(j_1,k_1)$ takes the following form:
\beqn
0\rar M_{j_1',k_1}\rar M(j_1,k_1) \rar L(j_1,k_1) \rar 0, \label{6.50}
\eeqn
where $l(j_1,j_1')=1$,
from which the following long exact sequence can be
induced
\beq
\cdots \rar \h{n}_{rel}(M(j_1,k_1)\otimes L_2) \rar \h{n}_{rel}(L(j_1,k_1)
\otimes L_2) \rar \h{n+1}_{rel}(M_{j'_1,k_1} \otimes L_2) \rar \no \\
\rar \h{n+1}_{rel}(M(j_1,k_1) \otimes L_2) \rar \h{n+1}_{rel}(L(j_1,k_1)
\otimes L_2) \rar \h{n+2}_{rel}(M_{j'_1,k_1} \otimes L_2) \rar
\cdots,
\eeq
where $L_2=L(j_2,k_2)$. Use theorem~\ref{tm07} and note that there exists a
``dual relation''\cite{FY1} between
the structure of Verma modules at level $k$ and level $-k-3$, The results
are easily obtained.\\
($i\!v$) is the most complicated case, however the proof can go on as that
of minimal model in the Virasoro algebra \cite{LZ5}.
Without loss of generality, we assume that $k_1+3/2>0$, then $L(j_1,k_1)$
is a resolution of the following complex
\beqn
\cdots \stackrel{\partial_{i-1}}{\rar} M^{i}
\stackrel{\partial_i}{\rar}
M^{i+1}\stackrel{\partial_{i+1}}{\rar} \cdots
\stackrel{\partial_{-2}}{\rar}
M^{-1} \stackrel{\partial_{-1}}{\rar} M^0=M(j_1,k_1)
{\rar} 0,
\eeqn
where $M^i=M_{j_1^{i},k_1}\oplus M_{j_1^{-i},k_1}, \ l(j_1^{\pm i},j_1)=i$.
Then we get
\beqn
\h{n}_{rel}(L(j_1,k_1)\otimes L(j_2,k_2))=\h{n}_{rel}(H^0(M^i,\df_i)
\otimes L(j_2,k_2)).\label{s361}
\eeqn
If $(-1/2-j_2,k_1)\not\rar (j_1,k_1)$, then by the dual relation,
$(-1/2-j_1,k_2)\not\rar (j_2,k_2)$,
 moreover $(-1/2-j_1^{\pm i},k_2)\not\rar (j_2,k_2)$. By theorem \ref{tm07},
$\h{n}_{rel}(M^i\otimes L(j_1,k_1))\cong\emptyset$, so by the corollary
\ref{c03},
$\h{n}_{rel}(L(j_1,k_1)\otimes L(j_2,k_2)\cong\emptyset$.\\
If $(-1/2-j_2,k_1)\rar (j_1,k_1)$, it is easily seen that the double cohomology
in r.h.s.\ of eq.(\ref{s361}) both collapse at the second term (cf.\
\cite{LZ5}) by using theorem \ref{tm07}. So by theorem \ref{tm02}, we
have
\beq
\h{n}_{rel}(L(j_1,k_1)\otimes L(j_2,k_2))\cong \sum_{p+q=n}
H^p(\h{q}_{rel}(M^p\otimes L(j_2,k_2)),\df_p),
\eeq
from which we get eq.(\ref{6.90}).
This completes the proof of the theorem.

Even though we get some insight by calculating the BRST cohomology with
coefficient in various $\os$-modules, we have not yet reached the goal of
our present paper, i.e. to find out the physical space of the $\OSP/\OSP$
theory.
The reason is that in the calculation of the BRST cohomology
$H(V_1\otimes V_2\otimes gh,Q)$, usually we take $V_1$ to be the admissible
representation of $\cal G$, of which there are finite number inside the
``conformal grid". However, as discussed by many authors,
e.g.\ ref.\cite{GK,Is}, $V_2$ is considered to be a representation of
$G^{\bf C}/G$, and  the space of state can be decomposed into
a direct integral of the irreducible representations of the Kac-Moody
superalgebras.
It is natural to consider $V_2$ as Wakimoto modules which are the free field
realization of $G^{\bf C}/G$. In the remaining of this section, we shall
calculated the BRST cohomology with $V_1$ the admissible module
and $V_2$ the Wakimoto module.

But first we compute the BRST cohomology with coefficient in the Wakimoto
modules.
The matter part $W(j_1,k_1)$ is the Fock space realized in eq.(\ref{s404}),
and the Liouville part $W^*(j_2,k_2)$  in eq.(\ref{s405}).
And  $\td{k}+k+3=0,\ \td{\af}_+=\pm i\af_+$.
We write out the diagonal part of $\cal G$:
\beq
J_0^{tot,3}&=&c^+_0b_{+,0}-c^-_0b_{-,0}+\frac{1}{2}c^{\frac{1}{2}}_0b_{\frac{1}{2},0}
-\frac{1}{2}c^{-\frac{1}{2}}_0b_{-\frac{1}{2},0}+\nonumber \\
&
&\sum_{n<0}c^+_nb_{+,n}-c^-_nb_{-,n}+\frac{1}{2}c^{\frac{1}{2}}_nb_{\frac{1}{2},n}
-\frac{1}{2}c^{-\frac{1}{2}}_nb_{-\frac{1}{2},n}c^+_nb_{+,n}c^+_nb_{+,n}+J_0^3+\td{J}_0^3.\\
L_0&=&\sum_n:n(\r_{-n}\bt_n+\td{\r}_{-n}\td{\bt}_n+\psi_{-n}\psi^+_{-n}
+\td{\psi}_{-n}\td{\psi}^+_n+c^\af_{-n}b_{\af,n}):+\nonumber \\
&&\sum_{n>0}(\phi_{-n}\phi_n+\td{\phi}_{-n}\td{\phi}_n)+
\frac{1}{2}\phi_0(\phi_0+1/\af_+)+\frac{1}{2}\td{\phi}_0(\td{\phi}_0+1/
\td{\af_+})\no\\
&=&\hat{L}_0+\Delta_j+\Delta_{\td{j}},
\eeq
where
\beqn
\hat{L}_0=\sum_n:n(\r_{-n}\bt_n+\td{\r}_{-n}\td{\bt}_n+\psi_{-n}\psi^+_{-n}
+\td{\psi}_{-n}\td{\psi}^+_n+c^\af_{-n}b_{\af,n}):
+\sum_{n\neq0}:\phi^+_{-n}\phi_n
\eeqn
is the level operator.
Without loss of generality, we set $\td{\af}_+=i\af_+,\ \phi^{\pm}=
1/\sqrt{2}(\phi\pm i\td{\phi})$ and a degree on the free field
\beqn
\begin{array}{cccccccl}
deg:&c^\af_n,&\r,&\td{\r},&\psi,&\td{\psi},&\phi_n^+,&1\\
deg:&b_{\af,n},&\bt,&\td{\bt},&\psi^+,&\td{\psi}^+,&\phi_n^-,&-1
\end{array}
\eeqn
Then $\hat{Q}$ can be rewrite as $\hat{Q}=\hat{Q}_>+\hat{Q}_0$,
where the degree zero operator
\beqn
\hat{Q}_0=\sum_n-\frac{1}{2}c_n^-\bt_{-n}+\frac{1}{2}c^+_n\td{\bt}_{-n}+
1/4c_n^{\frac{1}{2}}\td{\psi}^+_{-n}
-1/4c_n^{-\frac{1}{2}}\psi^+_{-n}+\sum_{n\neq 0}\af_+/\sqrt{2}c_n^3\phi^-_{-n}
\eeqn
contains only quadratic terms with opposite degrees.

We need to calculate  $H(C^*_{rel},\hat{Q}_0)$. Note that
\beqn
\hat{L}_0=\{ \hat{Q}_0,\hat{G}^0_0\},
\eeqn
where
\beqn
\hat{G}^0_0=\sum_n n:(b_{-,n}\r_{-n}-b_{+,n}\td{\r}_{-n}
+2b_{\frac{1}{2},n}\td{\psi}_{-n}-2b_{-\frac{1}{2},n}\psi_{-n})
+\sum_n\sqrt{2}/\af_+b_{3,n}\phi^+_n.
\eeqn
So a nontrivial $\hat{Q}_0$ states must be annihilated by  $\hat{L}_0$.
 We can just restrict ourselves to the subspace which consists of only
zero-mode excitation. However because it is a subspace of the relative
 complex, $L_0=0,\ J^{tot,3}_0=0$, so the subspace is nonempty only
when $\Delta_j+\Delta_{\td{j}}=0$.

To make our discussion more simple, the following relation is very
useful
\beqn
\begin{array}{lllll}
\hat{n}_1&=&\bt_0\r_0-b_{-,0}c^-_0&=&\{\hat{Q}_0,-2b_{-,0}\r_0\};\\
\hat{n}_2&=&\td{\bt}_0\td{\r}_0-b_{+,0}c^+_0&=&
\{\hat{Q}_0,2b_{+,0}\td{\r}_0\};\\
\hat{n}_3&=&-\td{\psi}_0^+\td{\psi}_0+b_{\frac{1}{2},0}c^{\frac{1}{2}}_0
  &=&\{\hat{Q}_0,4b_{\frac{1}{2},0}\psi_0\}; \\
\hat{n}_4&=&-\psi^+_0\psi_0+b_{-\frac{1}{2},0}c^{-\frac{1}{2}}_0
&=&\{\hat{Q}_0,-4b_{-\frac{1}{2},0}\psi_0\};
\end{array}
\eeqn
Note that in the Fock space $\hat{n}_i$ is diagonalable,
so a nontrivial $\hat{Q}_0$
states must satisfies $\hat{n}_i=0$.

As discussed previously, we have to specify the Fock space vacuum.
In calculating $\h{*}_{rel}(V_1\otimes V_2,\widehat{Q}_0)$, we first consider
the case that both $V_1$ and $V_2$ are highest weight modules. The relevant
Fock space vacuum is specified by the requirement that
\beq
&&\beta_n|j, \tilde{j}\rangle=
\tilde{\gamma}_n|j, \tilde{j}\rangle=
\psi^+_n|j, \tilde{j}\rangle=
\tilde{\psi}_n|j, \tilde{j}\rangle=0, \ \ \ \ \ n\geq 0 \no\\
&&\gamma_m|j, \tilde{j}\rangle=
\tilde{\beta}_m|j, \tilde{j}\rangle=
\psi_m|j, \tilde{j}\rangle=
\tilde{\psi}^+_m|j, \tilde{j}\rangle=0, \ \ \ \ \ n\geq 0 \\
&&(\partial\phi)_0\beta_n|j, \tilde{j}\rangle=\frac{2j}{\alpha_+}
\tilde{\gamma}_n|j, \tilde{j}\rangle,\ \ \ \
(\partial\tilde{\phi})_0\beta_n|j,
\tilde{j}\rangle=-\frac{2j}{\tilde{\alpha}_+}
\tilde{\gamma}_n|j, \tilde{j}\rangle.\no
\eeq
For the ghost vacua, we consider here the following two cases.

(i)Ghost vacuum $|gh\rangle_0$ \\
Recall that $|gh\rangle_0$ is defined by
\beq
b_{\alpha, n} |gh\rangle_0=c^\alpha_m |gh\rangle_0=0, \ \ n\geq 0,\ m>0,
\eeq
so that $|gh\rangle_0$ is a $\os$ singlet.
The ghost number operator is defined by
\beq
J^{gh}_0=\sum_{n>0, \af}(-1)^{d(\af)+1}b_{\alpha,-n}c^\alpha_n +
\sum_{n\geq 0, \af}c^\alpha_{-n}b_{\alpha,n}.
\eeq
So we have $J^{gh}_0|gh\rangle_0=0$, i.e. $|gh\rangle_0$ has ghost number
zero.
Note that
\beqn
\begin{array}{ll}
n_1=n_\r+n_{c^-},&n_2=-n_{\td{\bt}}+n_{c^+}-1,\\
n_3=-n_{\td{\psi}}+n_{c^{\frac{1}{2}}}+1,&n_4=n_\psi+n_{c^{-\frac{1}{2}}}.
\end{array}
\eeqn
Keep in mind  that $\bt,\r,\td{\bt},\td{\r}$ are bosonic fields, while
$\psi,\psi^+,\td{\psi},\td{\psi}^+$ are fermionic fields, we get the only
solution for $\hat{n}_i=0$ is
\[ n_\r=n_{\td{\bt}}=n_{c^-}=n_{c^\frac{1}{2}}=n_{c^{-\frac{1}{2}}}
=n_\psi=0,\
n_{c^+}=n_{\td{\psi}^+}=1.
\]
Upon the constraint that $J^{tot,3}_0=L_0=0$,
the $\hat{Q}_0$ cohomology state is
\beqn
\td{\psi}_0^+c^+_0|j, \tilde{j}\rangle\otimes |gh\rangle_0,
\eeqn
when $j+\td{j}+\frac{1}{2}=0$. Otherwise the cohomology is empty.

($i\!i$)Ghost vacuum $|gh\rangle_{0, \frac{1}{2}}$. \\
Such ghost vacuum was used, e.g. in ref.\cite{LZ2} and is specified as follows,
\beq
b_{\alpha, n} |gh\rangle_{0,\frac{1}{2}}=
c^\alpha_n |gh\rangle_{0,\frac{1}{2}}=0, \ \ n>0,\no\\
c^+_0|gh\rangle_{0,\frac{1}{2}}= c^{\frac{1}{2}}_0|gh\rangle_{0,\frac{1}{2}}=0
\eeq
It is easy to verify that $|gh\rangle_{0, \frac{1}{2}}$ is an
isospin-$\frac{1}{2}$ $\os$ highest weight state. If we fermionize
the bosonic ghost as in ref.\cite{FMS} and consider the two ghost vacua at
the different Bose sea level, we will find that $|gh\rangle_{0, \frac{1}{2}}$
has ghost number zero.
Note that
\beqn
\begin{array}{ll}
n_1=n_\r+n_{c^-},&n_2=-n_{\td{\bt}}-n_{b_+},\\
n_3=-n_{\td{\psi}}-n_{b_{\frac{1}{2}}},&n_4=n_\psi+n_{c^{-\frac{1}{2}}}.
\end{array}
\eeqn
 The only
solution for $n_i=0$ is
\[ n_\r=n_{\td{\bt}}=n_{c^-}=n_{c^\frac{1}{2}}=n_{c^{-\frac{1}{2}}}
=n_\psi=n_{c^+}=n_{\td{\psi}^+}=0.
\]
Upon the condition that $J^{tot,3}_0=L_0=0$,
the $\hat{Q}_0$ cohomology state is
\beqn
|j, \tilde{j}\rangle\otimes |gh\rangle_{0,\frac{1}{2}},
\eeqn
 when $j+\td{j}+1/2=0$. Otherwise the cohomology is empty.

Combining the above analysis, we have the following theorem,
\btm
(i) For ghost vacuum $|gh\rangle_{0}$,
\beqn
\h{n+1}_{rel}(W(j_1,k)\otimes W^*(j_2,-3-k)) \cong
\delta_{n,0}\delta_{j_1+j_2+1/2,0}.
\eeqn
($i\! i$)For ghost vacuum $|gh\rangle_{0,\frac{1}{2}}$,
\beqn
\h{n}_{rel}(W(j_1,k)\otimes W^*(j_2,-3-k)) \cong
\delta_{n,0}\delta_{j_1+j_2+1/2,0}.
\eeqn
\etm

As in \cite{HY2}, we also consider the cohomology $\h{*}_{rel}(V_1\otimes
V_2,\widehat{Q}_0)$ in the Fock space, when one of, or both $V_1,\ V_2$
are lowest weight modules (LWM). The results are list in table 1.
When they both are HWM (or LWM), we have $\h{*}_{rel}(V_1\otimes
V_2,\widehat{Q}_0)\cong \h{*}_{rel}(V_1\otimes V_2,\widehat{Q})$.
However, when one is LWM, another is HWM,
we only have an embedding instead of an isomorphism.
\begin{table} \label{fi01}
\begin{center}
\begin{tabular}{|c|c|c|}\hline\hline
\multicolumn{3}{|c|}{Ghost Vacuum $|gh\rangle_{0}$}\\ \hline
$V_2(\td{j},\td{k})\backslash V_1(j,k)$&HWM&LWM\\ \hline
HWM&$\td{\psi}^+_0c_0^+|\rangle,\ j+\td{j}+1/2=0$&$\psi_0^+\td{\psi}_0^+
c_0^-c_0^+|\rangle,\ j+\td{j}=0$\\ \hline
LWM&$|\rangle,\ j+\td{j}=0$&$\psi_0^+
c_0^-|\rangle,\ j+\td{j}-1/2=0$\\ \hline
\multicolumn{3}{|c|}{Ghost Vacuum $|gh\rangle_{0,\frac{1}{2}}$}\\ \hline
$V_2(\td{j},\td{k}) \backslash V_1(j,k)$&HWM&LWM\\ \hline
HWM&$|\rangle,\ j+\td{j}+1/2=0$&$\psi^+_0c_0^-|\rangle,\ j+\td{j}=0$\\ \hline
LWM&$\td{\psi}_0b_{+,0}|\rangle,\ j+\td{j}=0$&$\psi_0^+\td{\psi}_0
c_0^-b_{+,0}|\rangle,\ j+\td{j}-1/2=0$\\ \hline\hline
\end{tabular}
\caption{$\h{*}_{rel}(V_1\otimes V_2,\ \widehat{Q}_0)$}
\end{center}
\end{table}

{}From table 1, we see that when switching from our first ghost vacuum
into the second ghost vacuum, there is a change of the ghost number
by $1$ for the BRST state. In the following discussion, we assume that the
ghost vacuum is the former.
\blm
($i$)$\h{n}_{rel}(L(j_1,k_1)\otimes W^*(j_2,k_2))\neq 0$ iff $W_{-1/2-j_2,k_1}$
appears in the resolution of $L(j_1,k_1)$ in terms of Wakimoto modules.\\
($i\!i$) for $W(j_1,k_1)$ not in the case $I\!I\!I_+(\pm)$, if $W_{-1/2-j_2}$
appears in the resolution of $L(j_1,k_1)$,  then
\beqn
\h{n+1}_{rel}(L(j_1,k_1)\otimes W^*(j_2,k_2))\cong \delta_{n,sign
(j_1+j_2+1/2)l(j_1,-1/2-j_2)}.
\eeqn
\elm
Proof.($i$) If $\{W^i,\df_{i}\}$ is a resolution of $L(j_1,k_1)$,
using corollary
1, we get
\beq
\h{n}_{rel}(L(j_,k_1)\otimes W^*(j_2,k_2))&\cong&
\h{n}_{rel}(H^0(W^i,\df_i)\otimes W^*(j_2,k_2))\no\\
&\cong&
H^{n-1}(\h{1}_{rel}(W^i\otimes W^*(j_2,k_2)),\df_i),
\label{6.60}
\eeq
from which ($i$) is easily to get.\\
($i\!i$) Note that if the condition of ($i\!i$) holds, then it is in
the degree of $sign(j_1+j_2+1/2)l(j_1,-1/2-j_2)$. From
eq.(\ref{6.60}) we get ($i\!i$).\\
 This completes the proof of the lemma.
\btm\label{tms406}
($i$) for $W(j_1,k_1)$ in the case $I$,
\beqn
\h{n+1}_{rel}(L(j_1,k_1)\otimes W^*(j_2,k_2))\cong {\cal C}\delta_{n,0}
\delta_{j_1+j_2+1/2,0}.
\eeqn
($i\!i$) for $W(j_1,k_1)$ in the case $I\!I_+$,
\beqn
\h{n+1}_{rel}(L(j_1,k_1)\otimes W^*(j_2,k_2))\cong
\left\{\begin{array}{ll} {\cal C}\delta_{n,0}, &\mbox{if } j_1+j_2+1/2=0;\\
{\cal C}(\delta_{n,0}+\delta_{n,1}),&\mbox{if }-1/2-j_2\rar j_1,\\
&\mbox{ and }l(j_1,-1/2-j_2)=1;\\
\emptyset,&\mbox{otherwise}. \end{array}\right.
\eeqn
($i\!i\!i$) for $W(j_1,k_1)$ in the case $I\!I_-,$
\beqn
\h{n+1}_{rel}(L(j_1,k_1)\otimes W^*(j_2,k_2))\cong
\left\{\begin{array}{ll} {\cal C}\delta_{n,0}, &\mbox{if } j_1+j_2+1/2=0;\\
{\cal C}(\delta_{n,0}+\delta_{n,-1}),&\mbox{if }-1/2-j_2\rar j_1,\\
&\mbox{ and }l(j_1,-1/2-j_2)=1;\\
\emptyset,&\mbox{otherwise}. \end{array}\right.
\eeqn
($i\!v$) for $L(j_1,k_1)$ the admissible representation of $\os$
(c.f. eq.(\ref{s509})),
 \beqn
\h{n+1}_{rel}(L(j_1,k_1)\otimes W^*(j_2,k_2))\cong
\left\{\begin{array}{ll} {\cal C}\delta_{n,sign(j_1+j_2+1/2)
l(j_1,-j_2-1/2)}, &\mbox{if }
-j_2-1/2\rar j_1;\\
\emptyset,&\mbox{otherwise}. \end{array}\right.
\eeqn
\etm
Proof. Combining  theorem \ref{tm05} and the above lemma we get the theorem.

As a summary we restrict the matter part to the admissible representation
(cf. eq.(\ref{s509})), the following is our main results obtained in the above
discussion.
\beq
\h{n+1}_{rel}(L_{m,s}\otimes M_{2l\td{q}+m,s})&\cong&\delta_{n,|2l+1|};\no\\
\h{n+1}_{rel}(L_{m,s}\otimes M_{2l\td{q}-m,s})&\cong&\delta_{n,|2l|};\no\\
\h{n+1}_{rel}(L_{m,s}\otimes W^*_{2l\td{q}+m,s})&\cong&\delta_{n,(2l+1)};\no\\
\h{n+1}_{rel}(L_{m,s}\otimes W^*_{2l\td{q}-m,s})&\cong&\delta_{n,2l};
\label{s385}\\
\h{n+1}_{rel}(L_{m,s}\otimes L_{2l\td{q}+m,s})&\cong&
\left\{\begin{array}{ll}
{\cal C}\oplus {\cal C},&\mbox{if } n\in\mbox{odd,  and }|n|<|2l+1|,\\
{\cal C}, &\mbox{if } n=\pm(2l+1),\\
\emptyset,&\mbox{otherwise;}
\end{array}\right.\no\\
\h{n+1}_{rel}(L_{m,s}\otimes L_{2l\td{q}-m,s})&\cong&
\left\{\begin{array}{ll}
{\cal C}\oplus {\cal C},&\mbox{if } n\in\mbox{even,  and }|n|<|2l|,\\
{\cal C}, &\mbox{if } n=\pm 2l,\\
\emptyset,&\mbox{otherwise,}
\end{array}\right.\no
\eeq
where the subscript stands for the HW with $4j_{m,s}+1=m-s(2k_i+3),~i=1,2,~
2k_1+3=\td{q}/q,~2k_2+3=-\td{q}/q$.

Now that we have got the relative cohomology, the absolute cohomology is easy
to get. As discussed in \cite{LZ3,BMP2,HY2}, we also have
\beqn
\h{n}_{abs}\cong \h{n}_{rel}\oplus \h{n-1}_{rel},
\eeqn
where the coefficient is as that in eq.(\ref{s385}).

The cohomologies with coefficient being LWM$\otimes$LWM can be very easily
obtained. From the involution of the $\os$, eq.(\ref{s426})
\beqn
\begin{array}{cl}
J_n^\pm\rar -J_n^\mp,&J_n^3\rar-J_n^3,\\
j_n^\pm\rar\pm j_n^\mp,&k\rar k,~~~d\rar d,\end{array}\no
\eeqn
we get the isomorphism between LWM and HWM
\beq
M(j,k)\rar \td{M}(-j,k),~~~L(j,k)\rar\td{L}(-j,k),~~~W(j,k)\rar\td{W}(-j,k).
\label{060701}
\eeq
Moreover the ghost space $gh^{0}$ is invariant under such involution of
$\os$.
So we have
\beq \h{n}(\td{V}_1(-j_1,k_1)\otimes \td{V}_2(-j_2,k_2)\otimes
gh^{0},Q)\cong \h{n}(V_1(j_1,k_1)\otimes V_2(j_2,k_2)\otimes gh^{0}
,Q),
\eeq
where $\td{V}_i$ are LWMs while $V_i$ are corresponding HWMs under
map (\ref{060701}).

The cohomologies with ghost vacuum $|gh\rangle_{0,\frac{1}{2}}$ is found to be
isomorphic to
that with ghost vacuum $|gh\rangle_{0,\frac{1}{2}}$
by a shift of ghost number 1, i.e. we have
\beq
\h{n+1}_{rel}(V_1\otimes V_2\otimes gh^{0},\hat{Q})\cong
\h{n}_{rel}(V_1\otimes V_2\otimes gh^{0,\frac{1}{2}},\hat{Q}).
\eeq

In section 3, we have established the isomorphism between the R type
and the NS type modules
explicitly, and between the R type  ghost space $gh^{R,(0,\frac{1}{2})}$
 and the NS type ghost space $gh^{NS,(1,1)}$.
The BRST operator can be identified by such
an isomorphism. So the BRST states for the NS type $\os$ can also be obtained
by such a relation,
\beq
\h{*,NS}(V_1^{NS}\otimes V_2^{NS}\otimes gh^{NS,(1,1)},Q)
)\cong\h{*,R}(V_1^{R}\otimes V_2^{R}\otimes gh^{R,(0,\frac{1}{2})},Q),
\eeq
where $V_i^{R}$ and $V_i^{NS}$ are isomorphic to each other,
as listed in eq.(\ref{s433}).
Notice that for the NS type $\os$, the ghost vacuum $|gh\rangle_{0}$
annihilated by $b_0^\af$ can be got
from $|gh\rangle_{(1,1)}$  by the action of $b_{+,0}$ on the latter one. Namely
\beq
|gh\rangle^{NS}_{0}=b_{+,0}|gh\rangle^{NS}_{(1,1)}.
\eeq
So we also have
\beq
\h{n+1}(V_1^{NS}\otimes V_2^{NS}\otimes gh^{0},Q)\cong
h^{n}(V_1^{NS}\otimes V_2^{NS}\otimes gh^{(1,1)},Q)
\eeq

\section{Conclusion and Speculations}
The quantization of 2d supergravity has been approached by many
authors through the Hamiltonian reduction of the supergroup
valued WZNW model. However, in this paper, we take a somewhat different
procedure. By considering the $\OSP/\OSP$ gauged WZNW model, the
superconformal matter, 2d supergravity, and the super-reparametrization
ghost are naturally put together to form a covariant topological
conformal field theory. It remains a technical problem to find an
economical way of calculating the correlation functions in our formalism,
which deserves  our future investigation \cite{FY3}.

The physical state space in 2D gravity has been worked out completely
for $c\leq1$ \cite{LZ3,LZ4,BMP1}. However the similar work on the 2D
supergravity is far from complete\cite{BMP4}. When the matter sector is in the
minimal series (in fact it is always the case ), the cohomology like
this $H^{\frac{\infty}{2}+*}(L(c,\Delta)\otimes F_{\xi,\eta})$
does not appear in the literature due to the difficulty in constructing the
Felder BRST cohomology for the super-Virasoro algebra. Having established
the equivalence between fermionic strings and $\OSP/\OSP$ WZNW field theory,
now we may come to the answers, though partly, of the problem
arising in
ref.\cite{LZ5}. From theorem \ref{tms406}, we get the following proposition,
which is analogous to the results in 2D gravity\cite{LZ3}.
\bpp
Let $\xi=i\frac{q+\td{q}}{2\sqrt{q\td{q}}}$,
\begin{enumerate}
\item if $\eta=i\frac{2dq\td{q}+mq-s\td{q}}{2\sqrt{q\td{q}}}$,
\beq
H^{\frac{\infty}{2}+n}_{rel}(L(c_{\td{q},q},\Delta_{m,s})\otimes
F_{\xi,\eta})\cong
\delta_{n,-2d}{\cal C};
\eeq
\item if $\eta=i\frac{2dq\td{q}-mq-s\td{q}}{2\sqrt{q\td{q}}}$,
\beq
H^{\frac{\infty}{2}+n}_{rel}(L(c_{\td{q},q},\Delta_{m,s}\otimes
F_{\xi,\eta})\cong
\delta_{n,1-2d}{\cal C};
\eeq
\end{enumerate}
where the BRST states are in the NS (R) sector when $m+s$ is even (odd).
\epp

The problem remains on how to choose the ghost vacuum state. In the presented
paper, we have analysed two kinds of ghost vacuum states,
and found that the corresponding
BRST states differ by ghost number one. (The method can be generalized to
the construction of other different ghost vacua).
The situation is similar to the so called ``picture changing" operation
in critical fermionic string theory\cite{FMS}.
It is promissing that in this way one could establish the exact
correspondence between
the correlation functions of the ``fermion vertex" in noncritical
fermionic string theory and their counter parts in $\OSP/\OSP$ theory
\cite{FY3}.

\bigskip
\noindent
{\bf Acknowledgement}:
We are grateful to H.L. Hu and J.L. Petersen for useful discussions
and suggestions.
This work is supported in part by the National Science Foundation
of China and the National Science Committee of China. One of us (JBF)
would like
to thank ICTP  for kind hOSPitality during the last stage of the present
work.

\appendix
\section{Proof of the Twisted Reduction formula}
We first list some notations:\\
category $\cal O$ is $\os$-module category such that for any $\os$-module
$V$ in $\cal O$ satisfies
\begin{enumerate}
\item  $V$ is $Z_2$ graded which is consistent with that of $\os$;
\item $V=\oplus_{\lambda\in P(V)} V_{\lambda}, \
\mbox{dim}(V_{\lambda})<\infty$;
\item $\exists \lambda_1,\lambda_2,\ldots, \lambda_n\ s.t.\ P(V)\subset
\cup^n_{i=1}\lambda_i+\Delta_-$.
\end{enumerate}
$(\af, n) >0 $ iff $n>0$ or $n=0,\ \af>0$, and vice versa,
where $\af, ~n$ correspond to the
isospin and the conformal weight of the generators.\\
Let $c_{\af}=h_{\af\bt}c^{\bt},\ b^{\af}=b_{\bt}h^{\bt\af}$,
\beq
Q_+=\sum_{(-\af,n)<0}c_{\af,n}J_{-n}^\af-\sum_{(-\af,n),(-\bt,m)<0}
1/2f^{\af\bt}_{~~~\r}c_{\af,n}c_{\bt,m}b_{-n-m}^{\r},\no\\
Q_-=\sum_{(-\af,n)>0}c_{\af,n}\td{J}_{-n}^\af-\sum_{(-\af,n),(-\bt,m)>0}
1/2f^{\af\bt}_{~~~\r}c_{\af,n}c_{\bt,m}b_{-n-m}^{\r}.
\eeq
It can be verified that $Q_-^2=Q_+^2=Q_-Q_++Q_+Q_-=0$.

Rewrite the ghost space as $gh^{0}=gh^0_+\otimes gh_-^{0}$,
where $gh^0_+$ is generated by $c_{\af,n}$s $(-\af,n)<0$ and
$gh_-^{0}$ by $b_n^{\af},\ n<0$ and $c_{\af,0}, \ \af<0$.
For ghost vacuum $|gh\rangle_{0,\frac{1}{2}}$, we have the following
decomposition,
$gh^{(0,\frac{1}{2})}=gh^{0,\frac{1}{2}}_+\otimes gh^{0,\frac{1}{2}}_-$,
where $gh^{0,\frac{1}{2}}_+$ is generated by $c_{\af,n}$s $(-\af,n)<0$
and
$gh^{0,\frac{1}{2}}_-$ generated by $b_n^{\af},\
(\af, n)<0$.

So let $V_+,\ V_-$ be modules in the category $\cal O$, then
$(V_+\otimes gh _+,Q_+)$ and $(V_-\otimes gh_-^{0},Q_-)$ are two
differential complexes. The following formula is helpful for our
proof of the twisted reduction formula.
\bpp[K\H{u}nneth formula]
\beq
H^n_{rel}(V_+\otimes V_-\otimes gh^{0},Q_-+Q_+)\cong\sum_{p+q=n}
H^p(V_+\otimes gh_+,Q_+)\otimes H^q(V_-\otimes gh_-^{0},Q_-)[0].
\label{a045}
\eeq
\epp

\blm
\beq
dim(H^n(M(j,k)\otimes gh_-^0,Q_-)=\delta_{n,1},\label{060702}
\eeq
and a representation of the nontrivial $Q_-$ state  is
\[ (j_0^-c_{-,0}+c_{-1/2,0})|\rangle. \]
\elm
Proof. we can define a contracting homotopy operator,
\beq
\bar{G}=\sum_{(\af,n)>0}\bar{J}_n^{\af}h_{\af\bt}b_{-n}^{\bt},
\eeq
where the operators $\bar{J}_n^{\af},\ (\af,n)>0$ are defined such that
\beq
[J_n^{\af},\bar{J}_m^{\bt}]&=&(-1)^{d(\af)d(\bt)+1}[\bar{J}_m^{\bt},
J_n^\af]=\no\\
&=&\left\{\begin{array}{ll}
f^{\af\bt}_{\ \ \ \r}\bar{J}^{\r}_{n+m},& \mbox{if} (\af+\bt,n+m)>0;\\
h^{\af\bt}n\delta_{n+m,0},& \mbox{if} (\af+\bt,n+m)=0;\\
0,& otherwise,
\end{array}\right.
\\
\bar{J}_n^{\af}|\rangle=0.\no
\eeq
It can be verified that
\beq
\{Q_-,\bar{G}\}=\bar{L}_0=\sum_{(-\af,n)>0}((-1)^{d(\af)}nb_{-n}^\af c_{\af,n}
+J_{-n}^\af h_{\af\bt}\bar{J}_n^{\bt}),
\eeq
which satisfies
\beq
[\bar{L}_0,J_{-n}^\af]&=&n J_{-n}^\af;\no\\
{[}\bar{L}_0,b_{-n}^\af]&=&n b_{-n}^\af;\no\\
{[}\bar{L}_0,c_{\af,n}]&=&-n c_{\af,n};\\
{[}\bar{L}_0,Q_-]&=&0\no
 \eeq
This proves that the nontrivial $Q_-$  state should be $\bar{L}_0$ vanishing,
i.e. no negative mode excitations. Now by direct computation on the subspace
\[\{(j_0^-)^m(c_{-1/2,0})^n c_{-,0}|\rangle,
(j_0^-)^m(c_{-1/2,0})^n |\rangle\},\]
we get the unique nontrivial $Q_-$ states
\[(j_0^-c_{-,0}+c_{-1/2,0})|\rangle\]
up to  $Q_-$ exact states.\\
This completes the proof of the lemma.

Now by using the above proposition and lemma , we come to the proof of the
twisted reduction formula. It is analogous to the proof of the reduction
formula in ref.\cite{LZ2} by using $f$ degree on $V_+\otimes V_-\otimes
gh^{0}$. More explicitly we can define
\beq
&fdeg(J_n^\af)=-fdeg(\td{J}_n^\af)=-3n-2\af,\no\\
&fdeg(c_n^\af)=-fdeg(b_n^\af)=|3n+2\af|,
\eeq
then $\hat{Q}=\hat{Q}_0+\hat{Q}_>$, where $\hat{Q}_0=Q_++Q_-$.
Noticing  that the subspace of the relative complex with fixed
ghost number is finite dimensional, we get
\beq
\h{*}_{rel}(V_+\otimes V_-\otimes gh^0, \hat{Q})\cong
\h{*}_{rel}(V_+\otimes V_-\otimes gh^0, \hat{Q}_0),
\eeq
which together with eqs.(\ref{a045},\ref{060702}) gives the twisted reduction
formula eq.(\ref{061501}).

To compare the difference of the homology groups brought about by the different
ghost vacua, we list the theorem of homotopy Lie superalgebra, theorem
2.1 in ref.\cite{LZ2}
\begin{it}
\beq
H^n(M(\Lambda)\otimes gh_-,Q_-)\cong \delta_{n,0}{\cal C}[|\rangle].
\label{060703}
\eeq \end{it}
The difference between eq.(\ref{060702}) and eq.(\ref{060703}) leads also
the difference between the reduction formula \cite{LZ2} and the twisted one.

We have a conjecture here that eq.(\ref{060702}) can be generalized to
affine Lie superalgebra $\hat{G}$ associated with a finite dimensional
superalgebra $\cal G$
\beq
dim(H^n(M(\Lambda)\otimes gh_-^0,Q_-)=\delta_{n,
dim({\cal G}_0^+)},
\eeq
where $\cal G$ is the zero mode of $\hat{G}$, and ${\cal G}_0^+$ is Borel
even part of $\cal G$; $gh_-^0$ is generated by $b_n^\af$s with $n<0$ and
$c_{\bt,0}$ with $\bt$ being a negative root of $\cal G$.

\end{document}